\documentclass[aoas,preprint]{imsart}

\RequirePackage{amsthm,amsmath,amsfonts,amssymb}
\RequirePackage[authoryear]{natbib}
\RequirePackage{graphicx}

\usepackage{float}
\usepackage{capt-of}
\usepackage{booktabs}
\usepackage{multirow}
\usepackage{algorithm}
\RequirePackage[colorlinks,citecolor=blue,urlcolor=blue]{hyperref}
\usepackage{algpseudocode}
\usepackage[noabbrev,capitalise]{cleveref}

\makeatletter
\def\theHALG@line{\thealgorithm.\arabic{ALG@line}}
\makeatother

\DeclareMathSizes{11}{11}{7.6}{6}
\DeclareFontShape{T1}{ptm}{m}{scit}{<->ssub*ptm/m/it}{}

\begin{document}

\begin{frontmatter}

\title{Dual-Phase Flow Matching for Probabilistic Source Detection and Measurement in Astronomical Images}
\runtitle{Flow Matching for LSST}

\begin{aug}
\author[A]{\fnms{Yicun}~\snm{Duan}\ead[label=e1]{pduan@umich.edu}}
\and
\author[A]{\fnms{Jeffrey}~\snm{Regier}\ead[label=e2]{regier@umich.edu}\orcid{0000-0002-1472-5235}}
\address[A]{Department of Statistics, University of Michigan\printead[presep={ ,\ }]{e1,e2}}
\end{aug}

\begin{abstract}
Using the largest digital camera ever built, the NSF-DOE Vera C. Rubin Observatory's 10-year Legacy Survey of Space and Time (LSST) will image over half the night sky. Most astronomical analyses of LSST data operate not on images but on catalogs derived from them: tables of light sources detailing their positions, fluxes (colors and brightnesses), and types (e.g., star, galaxy). At LSST's imaging depths, light sources often overlap visually, so the configuration of sources in each image is itself uncertain. Nevertheless, LSST's cataloging pipeline commits to a single source configuration, which can bias downstream analyses.

Here, we instead approximate a posterior distribution over source configurations. Our posterior approximation is a continuous-time Markov process (CTMP) fitted by generator matching, an extension of flow matching to non-Euclidean state spaces. This CTMP generates a source configuration in two phases: it detects light sources in the first and measures them in the second. On simulated LSST-like data, the proposed method produces better posterior approximations than five comparison methods: two Markov chain Monte Carlo methods that run thousands of times more slowly at survey scales; two neural posterior estimation methods; and one ablation of the proposed method that lacks the dual-phase structure. Applied to LSST Data Preview 1 (DP1), the first public release of Rubin imaging, the proposed method alone recovers small-scale galaxy clustering in agreement with higher-resolution Hubble imaging and reports uncertainty that remains calibrated against that reference as blending increases. We deliver a probabilistic catalog of the Extended Chandra Deep Field South region, one of the most extensively studied extragalactic fields, and use it to characterize the galaxy color-magnitude distribution.
\end{abstract}


\end{frontmatter}

\section{Introduction}
\label{sec:introduction}

The NSF-DOE Vera C. Rubin Observatory's Legacy Survey of Space and Time (LSST) is a 10-year imaging program covering the night sky visible in the Southern Hemisphere in six broad photometric (wavelength) bands \citep{ivezic2019lsst}. Over its lifetime, the survey will collect 500 petabytes of imaging data and detect tens of billions of stars and galaxies. These data will support four primary scientific goals: probing dark energy and dark matter, exploring time-varying astronomical phenomena, mapping the Milky Way, and creating an inventory of the solar system. Most of the scientific analyses that pursue these goals do not directly interrogate the images. Instead, they operate on an astronomical catalog: a database of light sources inferred from survey images, with their positions, per-band brightnesses, morphologies, and classifications.

\begin{figure}[tb]
    \centering
    \includegraphics[width=\linewidth]{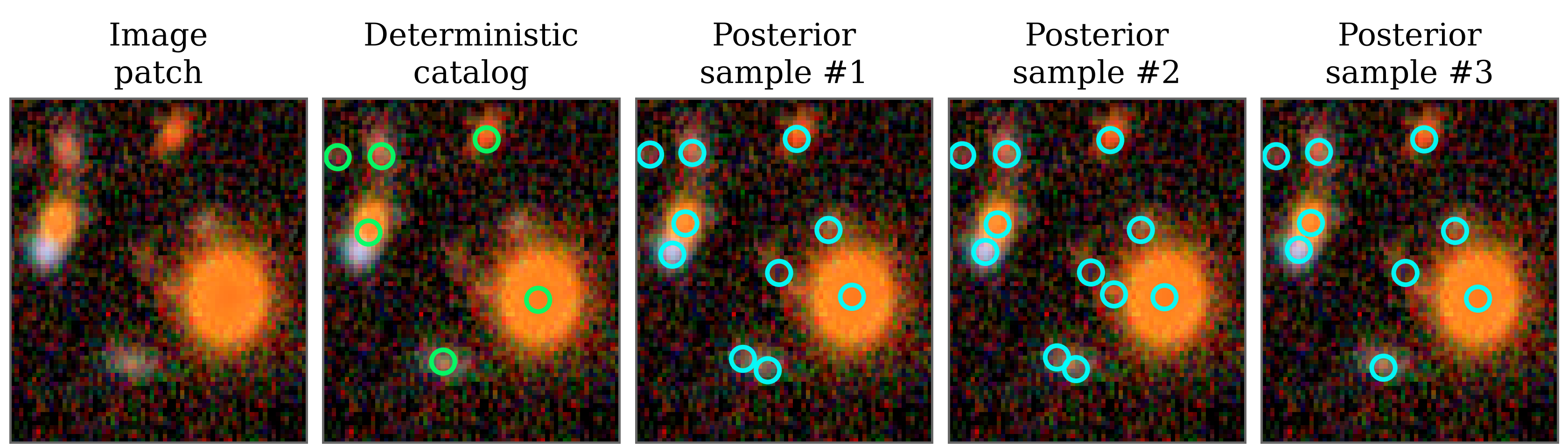}
    \caption{The cataloging task and two ways to solve it. \textbf{First panel:} an image patch with overlapping sources at LSST depth. \textbf{Second panel:} a deterministic catalog commits to a single answer. \textbf{Third to fifth panels:} three posterior samples from a probabilistic catalog, in which sources whose existence or position is individually uncertain appear in different places across samples.}
    \label{fig:cataloging_deterministic_vs_posterior_samples}
\end{figure}

Constructing an astronomical catalog from imaging data is an ill-posed inverse problem: many catalogs describing the light sources in a particular image are plausible. An image records only photoelectron counts, so the number of sources present, where each lies, and each source's brightness are latent. Near the survey's detection limit, where most sources reside, even the existence of a source is ambiguous. Despite this, the dominant cataloging methods today commit to a single configuration of sources. These methods are algorithmic pipelines in which each image is processed through stages that include detecting sources as peaks above a threshold in the image convolved with the point-spread function (a matched filter), associating detections of the same source across photometric bands, separating overlapping sources, and measuring each in turn \citep{bertin1996sextractor,lupton2001sdss,morganson2018dark}. Each stage commits to a single answer, which later stages treat as fixed; the catalog records as fact what the data leave undetermined (\Cref{fig:cataloging_deterministic_vs_posterior_samples}).
The LSST Science Pipelines, which will generate LSST's production catalogs \citep{bosch2018overview,rubin2025lsst}, and which we hereafter refer to collectively as the Rubin pipeline, largely adhere to this standard approach.

LSST will reach depths unprecedented for a wide-field, ground-based survey.
At these image depths, however, source ambiguity is the rule rather than the exception. Imaged galaxies are so dense that they often overlap on the focal plane. This overlap, called blending, has become a central obstacle in many cosmological analyses and is expected to rank among the primary sources of uncertainty for the next generation of sky surveys \citep{melchior2021challenge}. Blends come in two varieties: recognized and unrecognized. In a recognized blend, the overlapping galaxies are still detected as separate sources. In an unrecognized blend, the galaxies lie so close that traditional pipelines detect them as a single object, and the catalog does not indicate that anything was missed. At full LSST depth, roughly half of all galaxies are blended \citep{sanchez2021effects}, and unrecognized blends account for 14\% to 30\% of galaxies detected \citep{dawson2016ellipticity,troxel2023joint}.

Blending biases each of the major dark-energy probes, which infer cosmological parameters from statistics computed over the catalog \citep{mendoza2025blending}. \textit{Weak gravitational lensing} probes the mass distribution from small, spatially correlated distortions in galaxy shapes; blended source galaxies often go undetected, and the shapes of those detected are biased, so blending is a leading source of error limiting weak-lensing calibration \citep{nourbakhsh2022galaxy,maccrann2022des}. \textit{Photometric redshifts}, the per-galaxy distances inferred from per-band brightnesses, shift when light from a blended neighbor contaminates a galaxy's colors, most severely in unrecognized blends, where the neighbor can lie at a very different true distance. \textit{Galaxy clustering} measures the excess probability of finding a pair of galaxies at a given separation, from which the growth of cosmic structure is inferred; errors in galaxy counts directly distort this signal \citep{levine2025galaxy}. The LSST Dark Energy Science Collaboration (DESC) accordingly sets systematic-error budgets for these analyses \citep{mandelbaum2021lsst}.

\paragraph*{Probabilistic cataloging} An alternative to deterministic cataloging is to infer a posterior distribution over catalogs, thereby carrying detection and deblending ambiguities forward into downstream analyses rather than resolving them by fiat \citep{hogg2010telescopes}. Because the number of light sources in a given image is itself unknown, the posterior ranges over catalogs of varying dimension: catalog inference is \emph{transdimensional}. Early probabilistic cataloging methods used Markov chain Monte Carlo (MCMC): \citet{brewer2013probabilistic} sampled catalogs of crowded stellar fields with a transdimensional sampler, and subsequent work extended the approach to overlapping X-ray sources \citep{jones2015disentangling,meyer2021ebascs}, multiband photometry \citep{feder2020multiband}, and parametric galaxy models \citep{buchanan2023markov}. MCMC samples the posterior faithfully in the asymptotic limit, but on practical time scales often suffers from poor mixing: each MCMC iteration is itself slow, evaluating the likelihood over many pixels, and many iterations are required for detection problems, which tend to have highly nonconvex likelihoods \citep{regier2019approximate,liu2023variational}.

Neural posterior estimation (NPE), a type of simulation-based inference (SBI), is an alternative to MCMC that trains a neural network to map an image to an approximate posterior. \citet{liu2023variational} introduced NPE for stellar cataloging, and \citet{regier2026neural} improved its calibration with a spatially autoregressive variational family. These amortized variational inference methods are highly scalable, but their posterior approximations are confined to parametric families with restricted shapes and limited spatial dependence.

\paragraph*{Flow matching for simulation-based inference} Some recent neural SBI methods avoid a parametric variational family altogether, instead learning a transformation from a simple base distribution to the posterior \citep{sharrock2024sequential,geffner2023compositional}. \citet{wildberger2023flow}, in particular, fit a continuous normalizing flow (CNF) \citep{chen2018neural,grathwohl2019ffjord} to approximate the posterior using flow matching \citep{lipman2023flow}, which involves velocity regression rather than integration. This approach, however, requires a state space that is fixed-dimensional and Euclidean, and catalogs are neither: the number of sources is unknown, and each source mixes discrete attributes (whether it is present, and whether it is a star or a galaxy) with continuous ones (its position and flux).

\paragraph*{Our contributions} We make two methodological contributions (\Cref{sec:dpfm_methodology}) and use them to construct well-calibrated probabilistic catalogs of images from the first LSST preview release (\Cref{sec:dp1_dataset,sec:dp1_application}).

Our first methodological contribution is to generalize the flow-matching-for-SBI method proposed by \citet{wildberger2023flow} to the mixed continuous--discrete latent spaces required for astronomical cataloging, in which the number of objects is also unknown. We combine discrete flow matching \citep{gat2024discrete} for the categorical components with continuous flow matching \citep{lipman2023flow} for the Euclidean ones, under the generator-matching formalism \citep{holderrieth2025generator}, in a single amortized model.

Our second methodological contribution is a dual-phase probability path tailored to detection--measurement problems, in which light sources' attributes evolve only for sources that are determined to exist. We split the probability path into a detection phase, which infers which pixels contain source centroids, and a measurement phase, which infers each detected source's type, sub-pixel position, and flux.

We refer to the resulting method as Dual-Phase Flow Matching (DPFM).
On simulated data, DPFM outperforms four published methods for probabilistic cataloging (\Cref{sec:simulated_data_validation}).
DPFM yields better-calibrated detection and flux posteriors than two published Monte Carlo methods for probabilistic cataloging: a fixed-dimensional Hamiltonian Monte Carlo sampler \citep{stone2023astrophot} and a reversible-jump sampler that infers the number of sources \citep{portillo2017improved,feder2020multiband}. DPFM runs thousands of times faster than these MCMC-based methods at survey scales. On the LSST DESC Data Challenge 2 (DC2), a fully synthetic end-to-end LSST simulation also used to validate the Rubin pipeline, the proposed method yields a posterior that is better calibrated and that assigns higher log-likelihood to the true catalog than two amortized NPE comparators. Additionally, on DC2, DPFM approximately recovers the two-point correlation function characterizing galaxy spatial clustering, in contrast to the Rubin pipeline.

Applied to the first release of LSST images (\Cref{sec:dp1_dataset}), DPFM produces a high-fidelity probabilistic catalog, which we use to analyze the color-magnitude relationships among galaxies (\Cref{sec:dp1_application}).
DPFM avoids two classes of errors made by the Rubin pipeline, both arising where galaxies overlap. Where sources are blended, the Rubin catalog fits some of them with unrealistically large sizes (half-light radii) and correspondingly overestimated fluxes; and it underestimates how strongly nearby galaxies cluster, suppressing the two-point correlation function at small separations. Comparisons with Hubble data at higher angular resolution \citep{whitmore2016hubble,haussler2007gems} further reveal that the Rubin pipeline's reported uncertainties lose coverage as blending increases, whereas DPFM's remain close to nominal coverage. The nonconvexity of the loss function optimized by the Rubin pipeline in its composite-model (cModel) stage may explain how blending produces the first, whereas DPFM, which marginalizes rather than minimizes, may have a more favorable optimization landscape. The chief limitation of DPFM is its dependence on an imperfect simulator (\Cref{sec:discussion}).

\section{LSST Data Preview 1}
\label{sec:dp1_dataset}
\begin{figure}[t]
    \centering
    \includegraphics[width=0.78\linewidth]{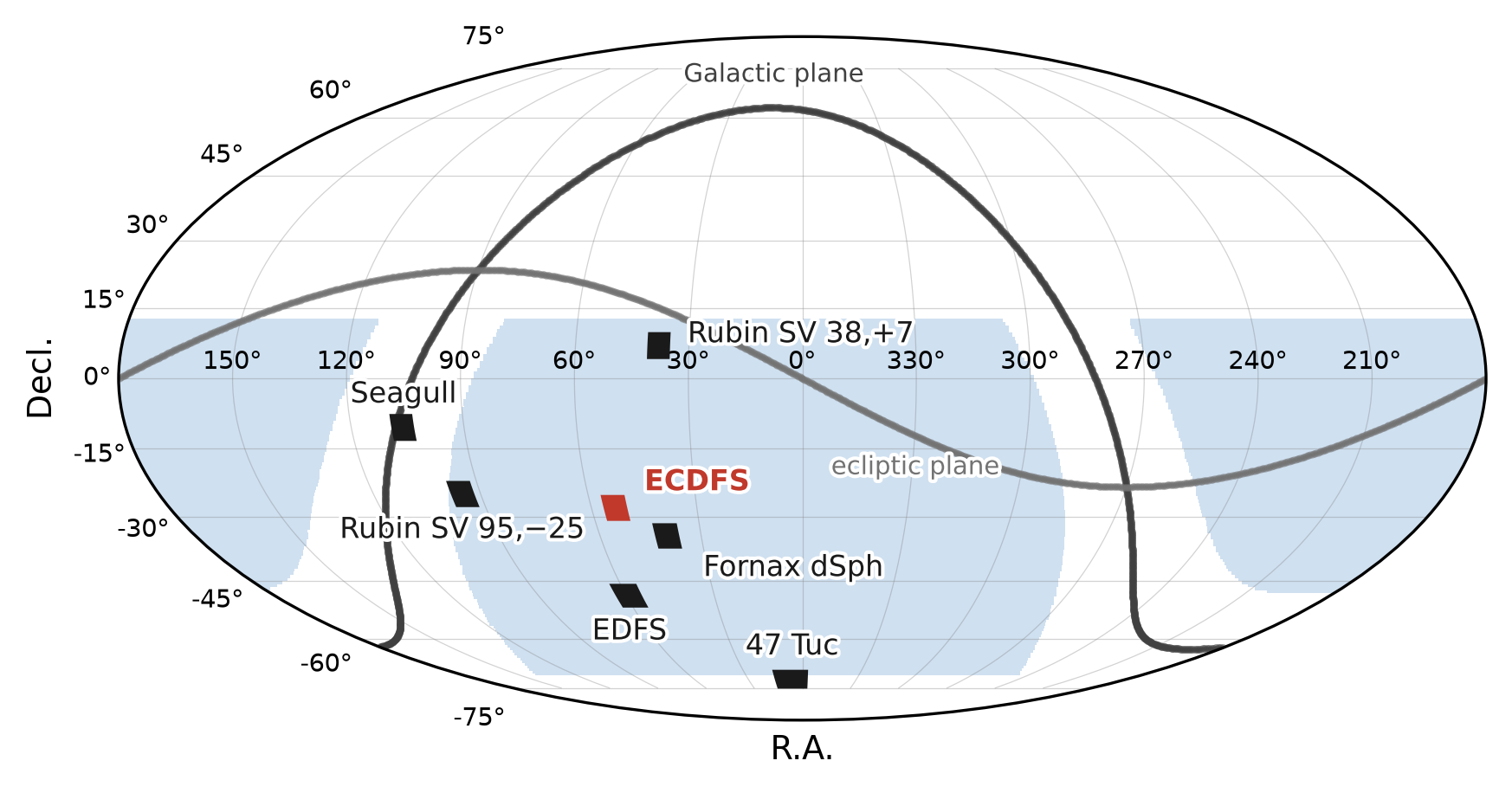}
    \caption{An all-sky, equal-area map showing the LSST survey footprint (shaded area). The seven DP1 fields are shown as labeled squares, enlarged for visibility. ECDFS, which we study, is the deepest of them.}
    \label{fig:dp1_fields_sky_map}
\end{figure}

LSST Data Preview~1 (DP1) is the Rubin Observatory's first release of on-sky imaging \citep{vera2026vera}, taken during a trial run with the LSST Commissioning Camera between October 24 and December 11, 2024. DP1 spans seven discontiguous, roughly equal-sized fields (\Cref{fig:dp1_fields_sky_map}), some 15 square degrees in all, approximately one-thousandth of the eventual LSST footprint.
The fields were chosen to be heterogeneous, so that the Rubin pipeline could be tested across a wide range of observing conditions. Of them, two are crowded stellar fields (the globular cluster 47 Tucanae and the Seagull Nebula), three sample varied stellar backgrounds (the Fornax Dwarf Spheroidal and two low-latitude science-validation fields), and two are extragalactic fields.

DP1 includes all six LSST broadband filters, $u$, $g$, $r$, $i$, $z$, and $y$, which span the near-ultraviolet through the near-infrared. The images are released as \emph{coadds}: images produced by stacking many short exposures of the same field \citep{rubin2025legacy}. By combining registered exposures on a common pixel grid, coadds offer greater photosensitivity (depth) than single-exposure images. The Rubin pipeline \citep{rubin2025lsst} assembles coadds by correcting every exposure for detector artifacts such as bias, crosstalk, and nonlinear response; subtracting the background; modeling the point-spread function (PSF); making an initial detection and measurement; calibrating the brightnesses to a common physical scale; and warping and stacking the exposures onto a shared grid. The pixel scale for LSST coadds is the same as for single exposures: 0.2 arcseconds per pixel side.

\begin{figure}[tb]
\centering
\begin{minipage}[t]{0.58\linewidth}\vspace{0pt}
    \centering
    \includegraphics[width=\linewidth]{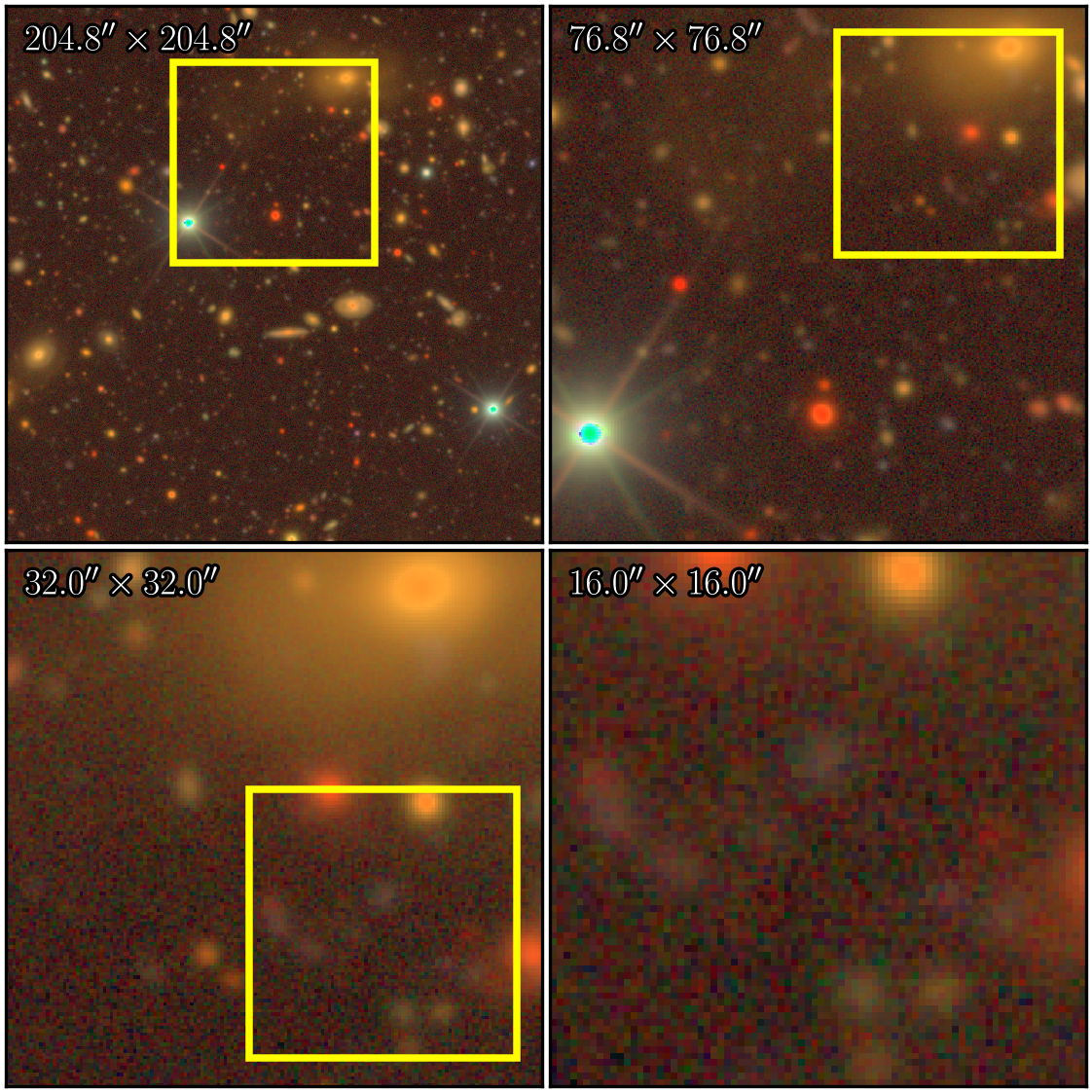}
    \caption{Example coadd cutouts from a DP1 field with progressive zoom. Each image is a cutout from the Extended Chandra Deep Field South region.}
    \label{fig:dp1_ecdfs_progressive_zoom}
\end{minipage}
\hfill
\begin{minipage}[t]{0.38\linewidth}\vspace{0pt}
    \centering
    \includegraphics[width=\linewidth]{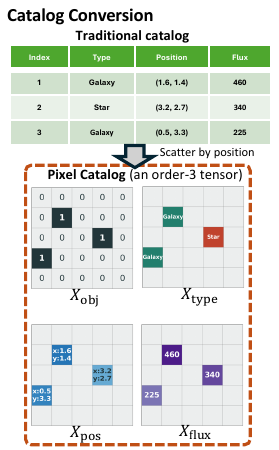}
    \caption{Catalog conversion. The traditional catalog is a table with a variable number of rows. We convert it to a pixel catalog via a positional scatter operation.}
    \label{fig:traditional_to_pixel_catalog}
\end{minipage}
\end{figure}

The Extended Chandra Deep Field South (ECDFS) region, the focus of our analysis, received 855 of DP1's 1,792 exposures and has the highest fraction of valid pixels, making it the deepest and most completely observed field in the release \citep{vera2026vera}. Its median seeing, the angular width to which the atmosphere blurs a point source, is 1.1 arcseconds; at this resolution, galaxies routinely overlap on the focal plane. Higher-resolution Hubble imaging provides independent reference data across part of ECDFS.
The ECDFS coadds reach a point-source depth of $r \approx 26$ over most of their footprint \citep{vera2026vera}, meaning that an isolated point source at that magnitude has a signal-to-noise ratio of five, the conventional reference point for comparing survey depth. (Smaller magnitudes are brighter, and one magnitude is a factor of about $2.5$ in flux.) That places ECDFS about one magnitude deeper than the Dark Energy Survey \citep{abbott2021dark,hartley2022dark}, which today is among the deepest wide-area optical surveys, but roughly a magnitude and a half short of LSST's 10-year depth \citep{ivezic2019lsst}. \Cref{fig:dp1_ecdfs_progressive_zoom} shows a progressive zoom into an ECDFS coadd.

The DP1 release also includes the Rubin pipeline's DP1 catalog, containing $2.3$ million light sources \citep{vera2026vera}. The Rubin pipeline is a production system built over more than a decade by a large collaboration \citep{bosch2018overview}, and its catalogs will serve as the interface through which nearly all LSST science is conducted. It detects sources above a threshold, deblends overlapping sources, and fits each source with a PSF-convolved model that yields its position, flux, and morphology; galaxy fluxes come from the pipeline's cModel fit, which combines exponential and de Vaucouleurs profiles \citep{bosch2018hsc}.
The Rubin catalog is the input to every analysis in the LSST dark-energy program, including weak-lensing shape measurement, photometric redshifts, and galaxy clustering. Any error in detecting or measuring a source propagates into each of them \citep{mandelbaum2021lsst}.
Until now, the pipeline has been characterized chiefly through simulations. DP1 is the first opportunity to assess the pipeline's cataloging on real Rubin data.

\section{Preliminaries}
\label{sec:preliminaries}

Our approach to cataloging combines several established techniques.
Throughout, uppercase letters denote random quantities and lowercase letters denote their realizations. We write $X_t$ for a time-varying random catalog and $x_t$ for a possible value, with $t \in [0,1]$.

\subsection{Catalog representation}
\label{sec:catalog_representation}

Astronomical catalogs are unordered, variable-size sets of light sources, which are traditionally represented as tables (\Cref{fig:traditional_to_pixel_catalog}, top).
The varying, unbounded number of rows in these tables, along with the lack of alignment between rows and image positions, makes tabular catalog representation difficult to work with. We instead adopt the pixel-catalog representation of \citet{liu2023variational} (\Cref{fig:traditional_to_pixel_catalog}, bottom).
In this representation, the catalog is a fixed-shape tensor with two spatial axes and one channel axis.
Light sources are spatially partitioned based on the positions of their centroids.
For each pixel, the catalog parameters indicate whether a source is centered in the pixel and, if so, its attributes.
If no light source is centered in a pixel, its channels are padded with zeros.

A pixel catalog represents a bounded number of sources per pixel.
In this work, we take the bound to be one: it is rare for a single pixel to contain the centroid of more than one source, and when one does, one source is typically much brighter than the other.
We restrict the support of our posterior approximations (but not of the forward model) to such pixel catalogs.
With this choice, inference over an unknown and potentially unbounded number of sources becomes inference over a fixed-shape, mixed discrete--continuous representation aligned with the image. A convolutional neural network can output this representation directly.

To define our pixel catalog more formally, let $H$ be the image height and $W$ be the image width, and let $B = 6$ be the number of photometric bands in LSST. Let brackets denote tensor concatenation along the channel axis, with a tensor of shape $H \times W$ treated as having one channel.
Then, a pixel catalog is $x = [x^\mathrm{occ}, x^\mathrm{type}, x^\mathrm{pos}, x^\mathrm{flux}]$.
Let $[\,\cdot\,]^i$ select the $i$-th pixel of a tensor (all channels at that pixel), where $i$ ranges over the $HW$ pixels.
The binary occupancy pattern $x^{\mathrm{occ}} \in \{0, 1\}^{H \times W}$ satisfies $[x^{\mathrm{occ}}]^i=1$ exactly when a source is centered in pixel $i$. At occupied pixels, $x^{\mathrm{type}} \in \{0, 1\}^{H \times W}$ records source types ($0$ for a star and $1$ for a galaxy), and $x^{\mathrm{pos}} \in \mathbb R^{2 \times H \times W}$ records sub-pixel positions, taking the center of the pixel as the origin; a source centroid's position is its integer-pixel coordinate plus this offset. The tensor $x^{\mathrm{flux}} \in \mathbb R^{B \times H \times W}$ records per-band standardized log-fluxes: a source with flux $f_b$ nanojanskys in band $b$ is recorded as $x^{\mathrm{flux}}_b = (\ln f_b - \mu_b)/\sigma_b$, where $\mu_b$ and $\sigma_b$ are fixed per-band constants.
At an unoccupied pixel, where $[x^{\mathrm{occ}}]^i=0$, we set $[x^{\mathrm{type}}]^i=[x^{\mathrm{pos}}]^i=[x^{\mathrm{flux}}]^i=0$. These zeros are padding rather than measured attributes.
We group channels by numerical type as $x^{\mathrm{disc}}:=[x^{\mathrm{occ}},x^{\mathrm{type}}]$ and $x^{\mathrm c}:=[x^{\mathrm{pos}},x^{\mathrm{flux}}]$, and the measurement channels as $x^{\mathrm m}:=[x^{\mathrm{type}},x^{\mathrm{pos}},x^{\mathrm{flux}}]$. Note that source type belongs to both the discrete channels and the measurement channels.

\subsection{Flow-matching posterior estimation}
\label{sec:flow_matching_posterior_estimation}

Let $Y$ denote a random image in the training population and $y$ an observed image on which inference is conditioned. NPE learns a conditional density estimator $q_\theta(x\mid y)\approx p(x\mid y)$ from simulated realizations of a random pair $(X,Y)$ with joint law $p$, where $\theta$ denotes the weights of a neural network \citep{papamakarios2016fast}. Once trained, the estimator amortizes inference across observations. Where the weights are not at issue, we write $q := q_\theta$.

\citet{wildberger2023flow} proposed flow-matching posterior estimation (FMPE), an NPE method that approximates the posterior with the terminal distribution of an observation-conditioned CNF. Flow matching trains its velocity field by regression on simulated samples, without integrating the flow during training. Under the assumptions of \citet[Thm.~1]{wildberger2023flow}, a constant times the square root of the flow-matching loss upper-bounds the expected forward Kullback--Leibler divergence, the criterion underlying NPE.

FMPE assumes that the latent state is a Euclidean vector. A pixel-catalog encoding does not have this form: it combines discrete existence and source-type variables with continuous positions and fluxes, requiring a generalization of FMPE to more complex state spaces.

\subsection{Continuous-time Markov processes}
\label{sec:continuous_time_markov_processes}

Let $(X_t)_{t \in [0,1]}$ be a continuous-time Markov process (CTMP) with marginal law $p_t$. A time-dependent linear operator $\mathcal L_t$ on its state space is said to generate this process if, for all bounded test functions $f$ that are smooth in the continuous coordinates,
\begin{equation*}
    \frac{d}{dt}\,\mathbb E[f(X_t)]
    = \mathbb E\!\left[(\mathcal L_t f)(X_t)\right].
\end{equation*}

On a mixed state space $\mathcal X=\mathcal X^{\mathrm{disc}}\times\mathbb R^d$, write $x = [x^{\mathrm{disc}},x^{\mathrm c}]$. A generator combining deterministic motion in the continuous coordinates with jumps in the discrete coordinates has the form
\begin{equation}
    (\mathcal L_t f)(x)
    =
    \nabla_{x^{\mathrm c}} f(x)^\top v_t(x)
    +
    \sum_{\substack{\zeta \in \mathcal X^{\mathrm{disc}}\\\zeta\ne x^{\mathrm{disc}}}}
    Q_t\!\left(x^{\mathrm{disc}},\zeta\mid x^{\mathrm c}\right)
    \left\{f([\zeta,x^{\mathrm c}])-f(x)\right\},
    \label{eq:mixed_generator}
\end{equation}
where $v_t:\mathcal X\to\mathbb R^d$ is a velocity field and $Q_t:\mathcal X^{\mathrm{disc}}\times\mathcal X^{\mathrm{disc}}\times\mathbb R^d\to\mathbb R$ is, for each fixed $x^{\mathrm c}$, a rate matrix in its first two arguments: its off-diagonal entries are nonnegative and its rows sum to zero.
On a purely Euclidean state space, only the first term remains, and the CTMP is a CNF. On a purely discrete state space, only the second term remains, and the process is a continuous-time Markov chain (CTMC).

Once a generator and an initial law $p_0$ are specified, samples are drawn by simulating the CTMP from $t=0$ to $t=1$.

\subsection{Generator matching}
\label{sec:generator_matching}

Generator matching learns the generator of a CTMP without integration during training \citep{holderrieth2025generator,lipman2024flow}. This CTMP has marginal law $(p_t)_{t\in[0,1]}$ that connects a tractable base law $p_0$ to a target distribution $p_1$ (in our work, the posterior), typically with $X_0\sim p_0$ drawn independently of $X_1\sim p_1$. On continuous coordinates, a common choice is a linear bridge:
\begin{equation}
    X_t^{\mathrm c}=tX_1^{\mathrm c}+(1-t)X_0^{\mathrm c}.    \label{eq:linear_bridge}
\end{equation}
For binary discrete coordinates, let $M_t$ be an array of independent $\mathrm{Bernoulli}(t)$ variables with the same shape as $X^{\mathrm{disc}}$. A commonly adopted path is
\begin{equation}
    X_t^{\mathrm{disc}}
    =
    M_t\odot X_1^{\mathrm{disc}}+(1-M_t)\odot X_0^{\mathrm{disc}},
    \label{eq:discrete_interpolation}
\end{equation}
where $\odot$ is elementwise multiplication.
Under this path, let $p_{t\mid x_1}$ be the conditional law of $X_t$ given $X_1 = x_1$.

On continuous coordinates, generator matching reduces to continuous flow matching \citep{lipman2023flow}. The conditional target velocity is $v_t(x_t\mid x_1) = (x_1^{\mathrm c}-x_t^{\mathrm c})/(1-t)$, which along the random bridge simplifies to $v_t(X_t\mid X_1)=X_1^{\mathrm c}-X_0^{\mathrm c}$. Its conditional expectation given the current state is the marginal target field in the generator,
\begin{equation*}
    v_t(x)
    =
    \mathbb E\!\left[v_t(X_t\mid X_1)\mid X_t=x\right].
\end{equation*}
A learned velocity field $v_t^\theta$ is trained by least-squares regression against samples of the conditional target:
\begin{equation*}
    L_\mathrm{FM}(\theta)
    =
    \int_0^1
    \mathbb E_{X_1\sim p_1,\;X_t\mid X_1\sim p_{t\mid X_1}}
    \left\|v_t^\theta(X_t)-v_t(X_t\mid X_1)\right\|^2
    \,dt.
\end{equation*}
Throughout, $\|\cdot\|$ is the Euclidean norm on flattened coordinates.

On discrete coordinates, generator matching reduces to discrete flow matching \citep{gat2024discrete}.
For a possible destination value $a\in\{0,1\}$, define the learned categorical endpoint probability
\begin{equation*}
    \widehat p_{1,j}^\theta(a\mid x_t)
    :=
    \mathbb P_\theta\!\left(X_{1,j}^{\mathrm{disc}}=a\mid X_t=x_t\right),
\end{equation*}
where $j$ indexes the discrete coordinates (flattened).
The endpoint probabilities are fitted by cross-entropy minimization.
The corresponding coordinate-level transition rate is
\begin{equation}
    \label{eq:discrete_transition_rates}
    Q_{t,j}^\theta(a\mid x_t)
    =
    \frac{\widehat p_{1,j}^\theta(a\mid x_t)-\mathbf 1\{a=x_{t,j}^{\mathrm{disc}}\}}{1-t}.
\end{equation}
Writing $x=[x^{\mathrm{disc}},x^{\mathrm c}]$ and using $-j$ to denote all discrete coordinates except $j$, these coordinate rates induce the full-state rate matrix of \Cref{eq:mixed_generator}:
\begin{equation*}
    Q_t^\theta\!\left(x^{\mathrm{disc}},\zeta\mid x^{\mathrm c}\right)
    =
    \sum_j
    \mathbf 1\!\left\{\zeta_{-j}=x_{-j}^{\mathrm{disc}}\right\}
    Q_{t,j}^\theta(\zeta_j\mid x).
\end{equation*}

Substituting $v_t^\theta$ and $Q_t^\theta$ for the velocity and transition-rate functions in \Cref{eq:mixed_generator} defines the learned generator.
Conditional generation is obtained by also supplying image $y$ to the learned generator.
We use the more familiar term ``flow matching'' rather than ``generator matching'' outside of this section.

\section{Methodology}
\label{sec:dpfm_methodology}

\begin{figure}[tb]
    \centering
    \includegraphics[width=0.9\linewidth]{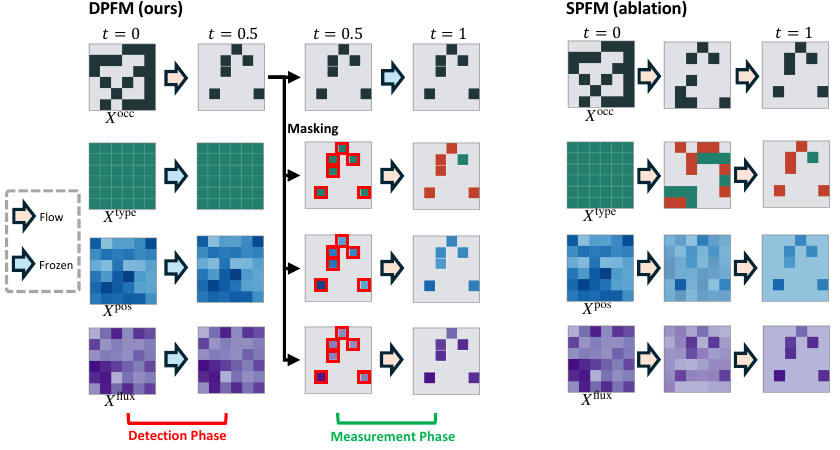}
    \caption{DPFM (the proposed method) and SPFM (an ablation of it) evolve the discrete--continuous state over the global sampling time $t \in [0,1]$. DPFM first evolves the occupancy pattern $X_t^{\mathrm{occ}}$ and then evolves source types, positions, and fluxes. SPFM evolves all components simultaneously.}
    \label{fig:dpfm_vs_spfm_flow_schedule}
\end{figure}

We propose Dual-Phase Flow Matching (DPFM), which fits a conditional density estimator $q_\theta(x \mid y)$ to approximate the posterior $p(x \mid y)$ over catalogs $x$ for any astronomical image $y$. We regard the random posterior catalog $X$ as the endpoint $X_1$ of a CTMP (\Cref{sec:continuous_time_markov_processes}), with realization $x_1$. Inference over a catalog separates into two questions: where are the sources, and what are they? A \emph{detection phase} infers the occupancy pattern $X_1^{\mathrm{occ}}$; a \emph{measurement phase} then infers source types, positions, and fluxes, collectively $X_1^{\mathrm m}=[X_1^{\mathrm{type}},X_1^{\mathrm{pos}},X_1^{\mathrm{flux}}]$, conditional on that resolved occupancy pattern. \Cref{fig:dpfm_vs_spfm_flow_schedule} shows these phases. The figure splits the global time interval $[0,1]$ between the two phases; below, each phase has its own $t\in[0,1]$. We now detail these phases in turn and then verify that collectively they target the posterior.

\paragraph*{Detection phase}
The detection phase targets $p(x^{\mathrm{occ}} \mid y)$, a marginal of the posterior. At $t=0$, the entries of $X_0^{\mathrm{occ}}$ are uniform on $\{0,1\}$, independently across pixels; at $t=1$, $X_1^{\mathrm{occ}}$ is the target occupancy pattern. We apply \Cref{eq:discrete_interpolation} independently to each entry $[X^{\mathrm{occ}}]^i$ to form the conditional probability path.
A neural network outputs the conditional endpoint probability $q_\theta([x_1^{\mathrm{occ}}]^i \mid x_t^{\mathrm{occ}},y)$ for each pixel $i$. Each of these probabilities conditions on the full current occupancy pattern, so a pixel's transition rate can depend on occupancy elsewhere. Substituting them for $\widehat p_{1,j}^\theta$ in \Cref{eq:discrete_transition_rates} gives the transition rates for the occupancy coordinates, and combining those per-pixel rates gives the rate matrix of \Cref{eq:mixed_generator}.
In the training losses below, each expectation includes a uniform average over $t\in[0,1]$ and an expectation over $(X_1,Y)$ with joint law $p$ and an independent base draw $X_0$, with $X_t$ given by the probability paths of \Cref{sec:generator_matching}. We train the detection network by minimizing the cross-entropy:
\begin{equation}
    L_\mathrm{detect}(\theta) = - \mathbb{E} \, \sum_{i=1}^{HW} \log q_\theta\!\left([X_1^{\mathrm{occ}}]^i \,\middle|\, X_t^{\mathrm{occ}},Y\right).
    \label{eq:detection_loss}
\end{equation}

\paragraph*{Measurement phase}
The measurement phase targets $p(x^{\mathrm m} \mid x^{\mathrm{occ}},y)$. We evaluate the measurement loss only where $[X_1^{\mathrm{occ}}]^i=1$, so the zero-padded attributes at unoccupied pixels never enter it. At these pixels, the base distribution is uniform on the discrete type and standard Gaussian on the continuous coordinates, $X_0^{\mathrm c} \sim \mathcal N(0, I)$. The continuous components follow the linear bridge of \Cref{eq:linear_bridge},
\begin{equation}
    X_t^{\mathrm c}
    =
    tX_1^{\mathrm c}+(1-t)X_0^{\mathrm c}
    \;\;\text{at pixels with } [X_1^{\mathrm{occ}}]^i=1,
    \qquad
    X_t^{\mathrm c}=[X_t^{\mathrm{pos}},X_t^{\mathrm{flux}}],
    \label{eq:occupied_pixel_continuous_interpolation}
\end{equation}
while the discrete source type follows \Cref{eq:discrete_interpolation} at each coordinate $[X^{\mathrm{type}}]^i$. The three measurement components follow their respective bridges at a common time and are inferred jointly, in lockstep, by one network with three measurement heads. The network takes as input the time $t$, the full noisy measurement state $x_t^{\mathrm m}$, the image $y$, and the occupancy pattern $x_1^{\mathrm{occ}}$. Its type head outputs the categorical endpoint probability $q_\theta([x_1^{\mathrm{type}}]^i \mid x_t^{\mathrm m},y,x_1^{\mathrm{occ}})$ for each pixel, and its position and flux heads output velocities $v_t^{\mathrm{pos},\theta}$ and $v_t^{\mathrm{flux},\theta}$, each evaluated at $(x_t^{\mathrm m},y,x_1^{\mathrm{occ}})$. We write $v_t^{\mathrm{pos},\theta,i}$ and $v_t^{\mathrm{flux},\theta,i}$ for their outputs at pixel $i$. Concatenating the two velocity fields gives $v_t^\theta:= [\,v_t^{\mathrm{pos},\theta},\;v_t^{\mathrm{flux},\theta}\,]$, the learned continuous field.

Each measurement head uses the objective of \Cref{sec:generator_matching} for its coordinate type (cross-entropy for the discrete type, and squared error against the random conditional target velocity $[X_1^{\mathrm c}]^i-[X_0^{\mathrm c}]^i$ for position and flux), all restricted to occupied pixels:
\begin{align}
    L_\mathrm{type}(\theta) &= - \mathbb{E} \, \sum_{i \,:\, [X_1^{\mathrm{occ}}]^i = 1} \log q_\theta\!\left([X_1^{\mathrm{type}}]^i \,\middle|\, X_t^{\mathrm m},Y,X_1^{\mathrm{occ}}\right), \label{eq:measurement_type_loss} \\
    L_\mathrm{pos}(\theta) &= \mathbb{E} \, \sum_{i \,:\, [X_1^{\mathrm{occ}}]^i = 1} \left\| v_t^{\mathrm{pos},\theta,i}(X_t^{\mathrm m},Y,X_1^{\mathrm{occ}}) - \left([X_1^{\mathrm{pos}}]^i - [X_0^{\mathrm{pos}}]^i\right) \right\|^2, \label{eq:measurement_position_loss} \\
    L_\mathrm{flux}(\theta) &= \mathbb{E} \, \sum_{i \,:\, [X_1^{\mathrm{occ}}]^i = 1} \left\| v_t^{\mathrm{flux},\theta,i}(X_t^{\mathrm m},Y,X_1^{\mathrm{occ}}) - \left([X_1^{\mathrm{flux}}]^i - [X_0^{\mathrm{flux}}]^i\right) \right\|^2. \label{eq:measurement_flux_loss}
\end{align}

The overall measurement loss combines the three per-head losses with learned weights, using the multitask objective proposed by \citet{kendall2018multi}:
\begin{equation*}
    L_\mathrm{measure}(\theta) = \sum_{k \in \{\mathrm{type},\, \mathrm{pos},\, \mathrm{flux}\}} \left\{ e^{-\lambda_k} L_k(\theta) + \lambda_k \right\},
\end{equation*}
where the $\lambda_k$ are per-head log-variances, learned jointly with the network weights.

\paragraph*{Training}
The full training loss is the sum of the two phases' losses, $L(\theta)=L_\mathrm{detect}(\theta)+L_\mathrm{measure}(\theta)$. The detection and measurement networks share no parameters, so we minimize the two terms in separate runs on simulated pairs $(x_1, y)$. Conditional on the ground-truth occupancy pattern $x_1^{\mathrm{occ}}$, the measurement network jointly minimizes its type, position, and flux terms, with every head conditioned on the full noisy measurement state.

\paragraph*{Sampling}
We draw each posterior sample in two phases. First, we initialize $x_0^{\mathrm{occ}}$ from its base law and simulate the learned CTMC to obtain $x_1^{\mathrm{occ}}$. Conditional on that realized occupancy pattern, we then initialize $x_0^{\mathrm m}$, solve the ODE $\tfrac{d}{dt} x_t^{\mathrm c} = v_t^\theta(x_t^{\mathrm m},y,x_1^{\mathrm{occ}})$, and simulate the CTMC for $x_t^{\mathrm{type}}$ to obtain $x_1^{\mathrm m}$.

The two phases therefore target the factors of the posterior,
\begin{equation*}
    p(x \mid y) = \underbrace{p(x^{\mathrm{occ}} \mid y)}_{\text{detection}} \cdot
    \underbrace{p(x^{\mathrm m} \mid x^{\mathrm{occ}},y)}_{\text{measurement}}.
\end{equation*}

Appendix A gives pseudocode for training and inference, and describes the network architecture, optimizer, schedule, sampling updates, and integration budget.

\subsection{Advantages of the dual-phase construction}
\label{sec:dual_phase_advantages}

To isolate the role of the dual-phase structure, we introduce an ablation having only a single phase (\Cref{fig:dpfm_vs_spfm_flow_schedule}).
We refer to this ablation as Single-Phase Flow Matching (SPFM).
SPFM jointly evolves detection and measurement: occupancy and source type start from uniform binary bases, and position and flux from standard Gaussian bases at every pixel; the attributes of unoccupied pixels end at the zero-padded catalog encoding.
Because SPFM performs inference this way, the zero-padded entries enter into the loss function at training time; the SPFM network must be trained to produce these zeros, which are abundant.
\Cref{sec:simulated_data_validation} compares the resulting performance of DPFM and SPFM.

\section{Validation on simulated data}
\label{sec:simulated_data_validation}
For simulated astronomical images, we know the ground-truth catalog, which facilitates validation.
We validate DPFM on two complementary simulated datasets before applying it to real Rubin imagery in \Cref{sec:dp1_application}. We compare against both MCMC-based and NPE-based comparators in terms of posterior calibration, recovery of spatial clustering structure, and inference speed.

\subsection{Simulated datasets}
\label{sec:simulated_datasets}

The first dataset, \emph{semi-synthetic DP1}, uses the positions, magnitudes, and morphologies reported in the Rubin DP1 catalog as the ground-truth catalog, ensuring a connection to the real data distribution. Images are generated with AstroPhot \citep{stone2023astrophot}, using a Moffat PSF and correlated Gaussian background noise, both of which are fitted to real DP1 images. Because one comparator (PCAT) models only stars, we also render a star-only version of each image, with sources' positions and fluxes and the image's depth and PSF held fixed. Appendix B gives further details.

The second dataset, the LSST DESC Data Challenge 2 Simulated Sky Survey \citep[\emph{DC2};][]{abolfathi2021lsst}, was built by the LSST Dark Energy Science Collaboration to prepare its dark-energy analyses for LSST data. A cosmological $N$-body simulation populates a wide sky patch with galaxies of known position, redshift, and morphology, and a Milky Way model supplies the stars; an instrument simulator renders them into LSST-like exposures in all six bands over five years of the planned cadence; and the Rubin pipeline reduces those to calibrated coadds and catalogs. Appendix C gives further details.

\begin{figure}[tb]
\centering

\begin{minipage}[b]{0.9\linewidth}
    \centering
    \begin{minipage}[b]{0.3\linewidth}
        \centering
        DP1\\[2pt]
        \includegraphics[width=\linewidth]{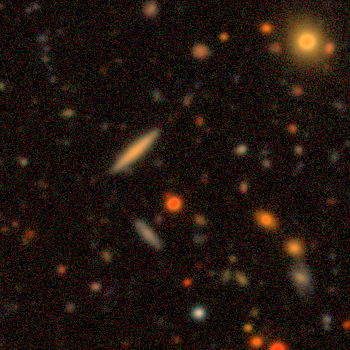}
    \end{minipage}
    \hspace{5pt}
    \begin{minipage}[b]{0.3\linewidth}
        \centering
        Semi-synthetic DP1\\[2pt]
        \includegraphics[width=\linewidth]{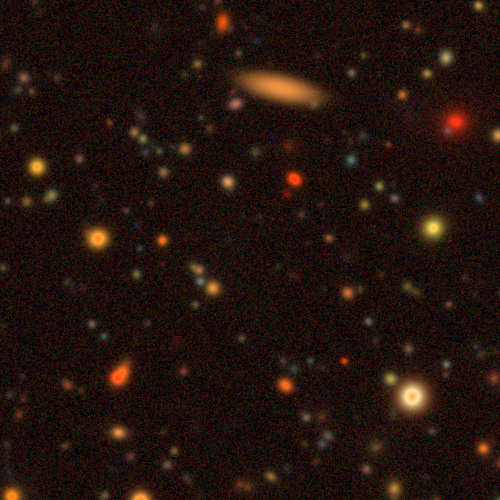}
    \end{minipage}
    \hspace{5pt}
    \begin{minipage}[b]{0.3\linewidth}
        \centering
        DC2\\[2pt]
        \includegraphics[width=\linewidth]{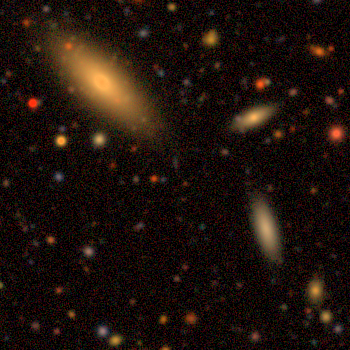}
    \end{minipage}
    \caption{Sample images from DP1 (real), semi-synthetic DP1, and DC2.}
    \label{fig:real_and_simulated_image_comparison}
\end{minipage}

\end{figure}

\Cref{fig:real_and_simulated_image_comparison} shows sample images from both datasets, alongside a real image from DP1.

\subsection{Comparison methods}
\label{sec:comparison_methods}

We compare the proposed method to four published comparators and SPFM, our ablation of the proposed method from \Cref{sec:dual_phase_advantages}.
The first two comparators are MCMC-based methods for probabilistic astronomical cataloging.
\emph{AstroPhot} runs a Hamiltonian Monte Carlo (HMC) sampler whose source count is fixed by peak-finding initialization \citep{stone2023astrophot}; it uses the exact correlated-Gaussian pixel-noise likelihood and the same PSF model used to generate semi-synthetic DP1 (Appendix B). \emph{PCAT} \citep{portillo2017improved,feder2020multiband} instead lets the source count vary, using reversible-jump MCMC with split-and-merge moves.

The next two are NPE methods, both of which restrict the posterior to a parametric variational family. \emph{Ind-NPE} \citep{liu2023variational}, extended to galaxies by \citet{mendoza2026simulation}, uses a mean-field factorization across image tiles; \emph{AR-NPE} \citep{regier2026neural}, further developed by \citet{duan2026neural}, conditions each tile on its neighbors to account for detection ambiguities at tile boundaries.

The MCMC comparators require likelihood evaluations, so we run them on semi-synthetic DP1, whose images we render ourselves; running them on DC2 would require a misspecified likelihood. The NPE comparators, SPFM, and DPFM need no likelihood evaluations, so we apply them to DC2, a simulation built and validated outside this work.

\begin{figure*}[tb]
    \centering
    \includegraphics[width=\textwidth]{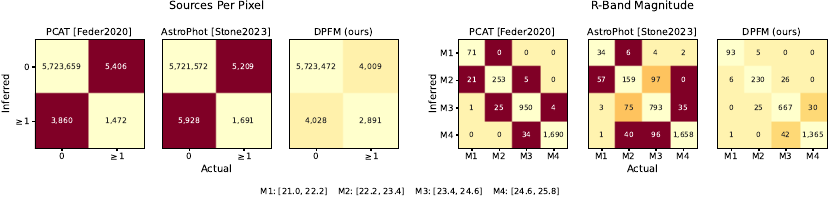}
    \caption{Confusion matrices of sources per pixel and $r$-band magnitude, comparing the proposed method with AstroPhot \citep{stone2023astrophot} and PCAT \citep{feder2020multiband} on semi-synthetic DP1. The matrix's symmetry measures the calibration of the posterior approximation. Cell colors show the asymmetry of each transpose pair: a cell and its reflection across the diagonal. Pairs whose counts are consistent with being equal, by an exact conditional test for equality of two Poisson means, are shown in light colors; the rest are darker the greater the asymmetry.}
    \label{fig:semi_synthetic_dp1_source_count_and_magnitude_confusion_matrices}
\end{figure*}

\subsection{Simulation-based posterior calibration diagnostics}
\label{sec:posterior_calibration_diagnostics}
Sampling the random pair $(X,Y)$ ancestrally from its joint law is equivalent to first sampling $Y$ from its marginal law and then $X$ from its posterior law conditional on $Y$, so simulation gives one exact posterior draw per data point. If $q(x \mid y)$ is accurate in aggregate over the marginal distribution of $Y$, the joint laws with densities $q(x \mid y)p(y)$ and $p(x \mid y)p(y)$ should be nearly indistinguishable. This is the idea behind simulation-based calibration (SBC) \citep{yao2018yes, talts2018validating}. Let $p_Y$ denote the marginal law of $Y$ and, conditional on $Y$, independently draw $X'\sim p(\,\cdot\mid Y)$ and $X''\sim q(\,\cdot\mid Y)$. For a summary statistic $T$ and measurable sets $\mathcal{I}$ and $\mathcal{J}$ in its range, our primary SBC diagnostic is the joint frequency with which the summary of a true-posterior draw lands in $\mathcal{I}$ and that of a variational draw in $\mathcal{J}$:
\begin{equation*}
C(\mathcal{I}, \mathcal{J})
=
\mathbb{E}_{\substack{Y\sim p_Y,\;X' \sim p(\,\cdot\mid Y),\;X'' \sim q(\,\cdot\mid Y)}}
\left[ \mathbf{1}\{ T(X') \in \mathcal{I} \wedge T(X'') \in \mathcal{J} \} \right],
\end{equation*}
where $\mathbf{1}\{\cdot\}$ is the indicator function. When $q$ matches $p$, the draws $X'$ and $X''$ are exchangeable, so $C$ is symmetric; hence, any departure from $C(\mathcal{I}, \mathcal{J}) = C(\mathcal{J}, \mathcal{I})$ measures miscalibration. We tabulate $C$ over a partition of the summary's range as a ``confusion matrix'' between true and inferred values. Concretely, calibration is read off each transpose pair of cells, cell $(a, b)$ against cell $(b, a)$; a well-calibrated posterior labels as many true-$a$ regions $b$ as it labels true-$b$ regions $a$.

We apply the diagnostic to four summary statistics: source count, sub-pixel position, per-band magnitude, and source type.
Source count is the number of sources in a small contiguous region.

\subsection{Semi-synthetic DP1 SBC results}
\label{sec:semi_synthetic_dp1_calibration}

On semi-synthetic DP1, we apply the MCMC-based comparators. For source detection (\Cref{fig:semi_synthetic_dp1_source_count_and_magnitude_confusion_matrices}, left), the confusion matrix is $2\times{}2$ and the transpose pair corresponds to false negatives versus false positives. For DPFM, this confusion matrix is nearly symmetric, with 4{,}009 false negatives against 4{,}028 false positives. Both MCMC comparators are highly imbalanced, with PCAT incurring disproportionately many false negatives and AstroPhot incurring disproportionately many false positives.

For flux estimation (\Cref{fig:semi_synthetic_dp1_source_count_and_magnitude_confusion_matrices}, right), both MCMC samplers systematically overestimate magnitudes, most severely in the brightest bin, where each scatters far more sources toward fainter bins than in the reverse direction. The proposed method's binned-magnitude confusion matrix is nearly symmetric across adjacent-bin transpose pairs. We gave MCMC a computational budget at the limit of our resources, and we chose the HMC and PCAT step sizes after testing a range of alternatives; the reported posteriors are the best we could achieve. \Cref{sec:runtime} provides details of the MCMC methods' runtime requirements.

\begin{figure*}[tb]
    \centering
    \includegraphics[width=0.82\textwidth]{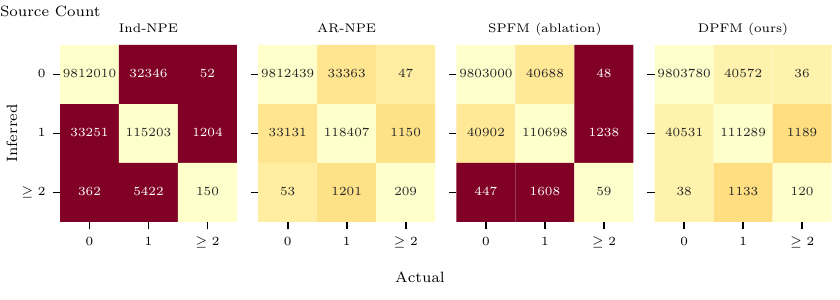}
    \includegraphics[width=0.82\textwidth]{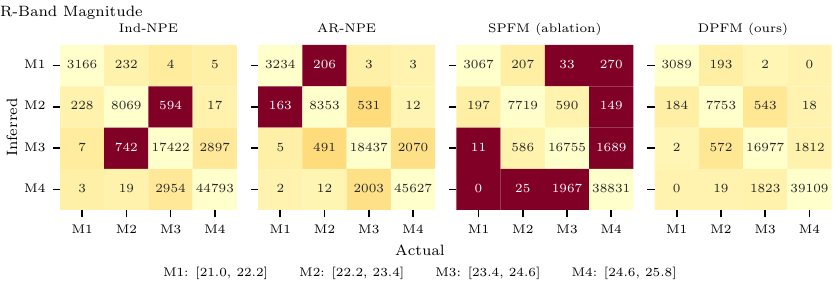}
    \includegraphics[width=0.82\textwidth]{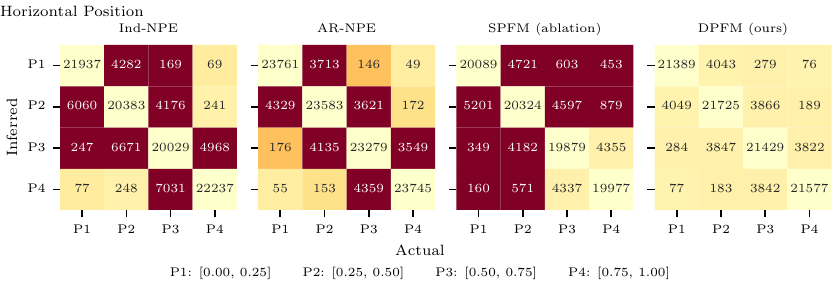}
    \caption{Confusion matrices for source count, $r$-band magnitude, and horizontal sub-pixel position on DC2, comparing the proposed method with the Ind-NPE \citep{liu2023variational} and AR-NPE \citep{regier2026neural} comparators, as well as the SPFM ablation. Cell colors follow the convention of \Cref{fig:semi_synthetic_dp1_source_count_and_magnitude_confusion_matrices}.}
    \label{fig:dc2_source_count_and_magnitude_confusion_matrices}
\end{figure*}

\subsection{DC2 SBC results}
\label{sec:dc2_calibration}

On DC2, we compare with the NPE comparators; all are likelihood-free, as are our diagnostics. \Cref{fig:dc2_source_count_and_magnitude_confusion_matrices} (top row) shows that the proposed method and the AR-NPE comparator yield symmetric confusion matrices for source count, the number of sources within a $4 \times 4$-pixel region. In contrast, the Ind-NPE comparator's confusion matrix is highly asymmetric. For example, between source counts one and two or more, it produces a transpose pair of 1{,}204 and 5{,}422, whereas the proposed method keeps that pair nearly balanced at 1{,}189 and 1{,}133.

For flux estimation, the proposed method yields the most symmetric confusion matrix (\Cref{fig:dc2_source_count_and_magnitude_confusion_matrices}, middle row). Sub-pixel position behaves the same way: DPFM is symmetric across adjacent-bin transpose pairs, while both NPE comparators are biased to the right, strongly for Ind-NPE and weakly for AR-NPE (\Cref{fig:dc2_source_count_and_magnitude_confusion_matrices}, bottom row).

The SPFM ablation's confusion matrices are asymmetric. In magnitude, SPFM labels the faintest sources as the brightest far more often than the reverse, whereas DPFM does not. In source count, SPFM remains balanced at counts zero and one and becomes more asymmetric at count two.

Appendix C provides further results, including source-type classification performance and the frequentist coverage of credible intervals.

\subsection{Held-out log-likelihood}

\begin{table}[t]
\centering
\small
\begin{tabular}{@{}lcccc@{}}
\toprule
 &
{\begin{tabular}[c]{@{}c@{}}Ind-NPE\\{[}Liu2023{]}\end{tabular}} &
{\begin{tabular}[c]{@{}c@{}}AR-NPE\\{[}Regier2026{]}\end{tabular}} &
{\begin{tabular}[c]{@{}c@{}}DPFM\\(ours)\end{tabular}} &
Difference \\
\midrule
Detection   & $-1093.7$ & $-1006.8$ & $\mathbf{-640.1}$ & $366.6 \pm 40.0$ \\
Measurement & $479.7$   & $681.8$   & $\mathbf{5272.0}$ & $4590.2 \pm 50.7$ \\
Overall     & $-614.0$  & $-325.0$  & $\mathbf{4631.9}$ & $4956.8 \pm 82.4$ \\
\bottomrule
\end{tabular}
\caption{Held-out log-likelihood on DC2, in nats per $512\times512$ image (higher is better).
``Detection'' is the log-likelihood that each posterior approximation assigns to the true occupancy pattern. ``Measurement'' is the log-likelihood it assigns to the true source types, sub-pixel positions, and fluxes conditional on the true occupancy pattern. ``Overall'' is their sum. Technically, the DPFM entries are lower bounds on its held-out log-likelihood. The ``Difference''
column is DPFM minus AR-NPE, with a standard error from a paired bootstrap over the 172 DC2 test images.}
\label{tab:dc2_held_out_log_likelihood}
\end{table}

As a proper scoring rule, log-likelihood simultaneously assesses the calibration and precision of posterior approximations.
Further, for NPE methods (Ind-NPE and AR-NPE), the log-likelihood is the training objective. By maximizing log-likelihood, NPE minimizes the expected Kullback--Leibler divergence from the posterior to its approximation. NPE does not target calibration or the other diagnostics separately; they are expected to hold insofar as this divergence is small. For DPFM, the held-out log-likelihood is an evaluation score rather than the training objective.

For Ind-NPE, AR-NPE, and DPFM, we compute the log-likelihood of the true catalog for 172 DC2 test-set images under
each method's posterior approximation. We evaluate Ind-NPE and AR-NPE at their native
$4\times4$-pixel tiling, with up to two detections per tile. We restrict the comparison to tiles whose true configuration every method can represent, excluding the $0.003\%$ of tiles that hold unresolved blends or more than two sources.

\Cref{tab:dc2_held_out_log_likelihood} shows that DPFM assigns a higher held-out log-likelihood than both NPE comparators, on both the detection and the measurement terms.
DPFM also assigns a higher log-likelihood to each test image's catalog individually.
The detection term reflects sharper estimates of the sparse occupancy pattern. The measurement term, a joint density over flux, source type, and sub-pixel position, accounts for most of the difference.

\subsection{Recovering the spatial clustering signal}
\label{sec:spatial_clustering}

\begin{figure}[tb]
    \centering
    \includegraphics[width=0.40\textwidth]{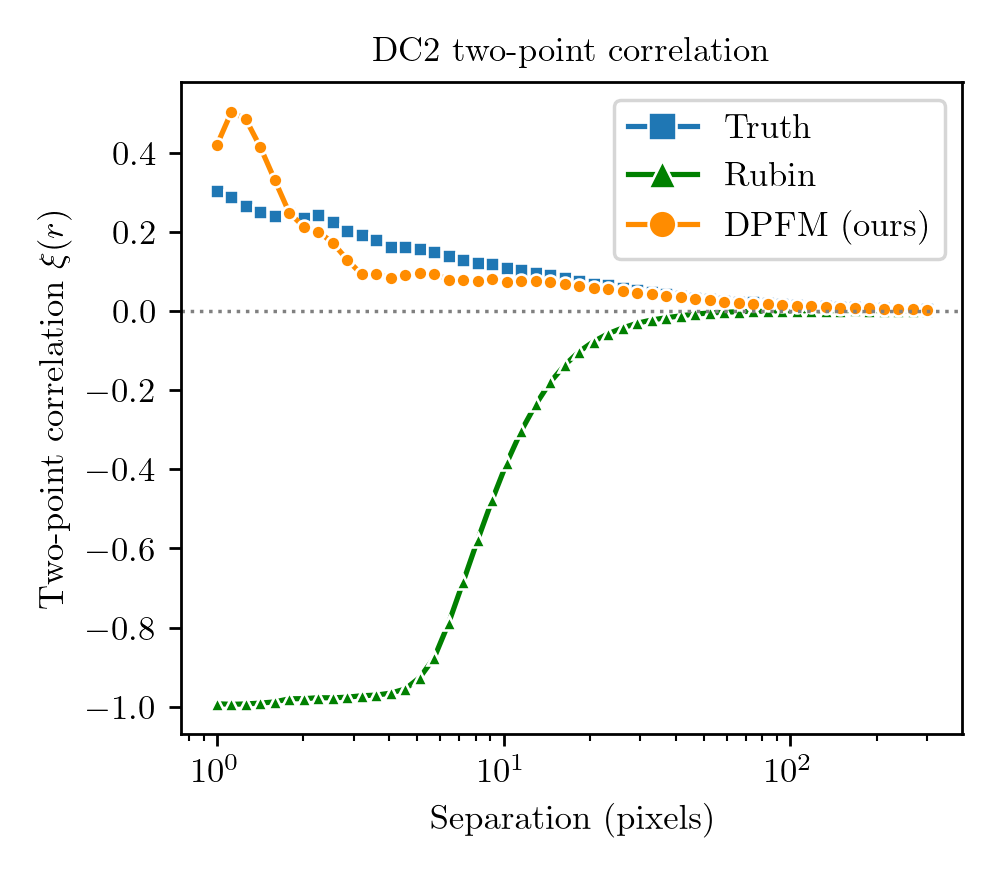}
    \includegraphics[width=0.40\textwidth]{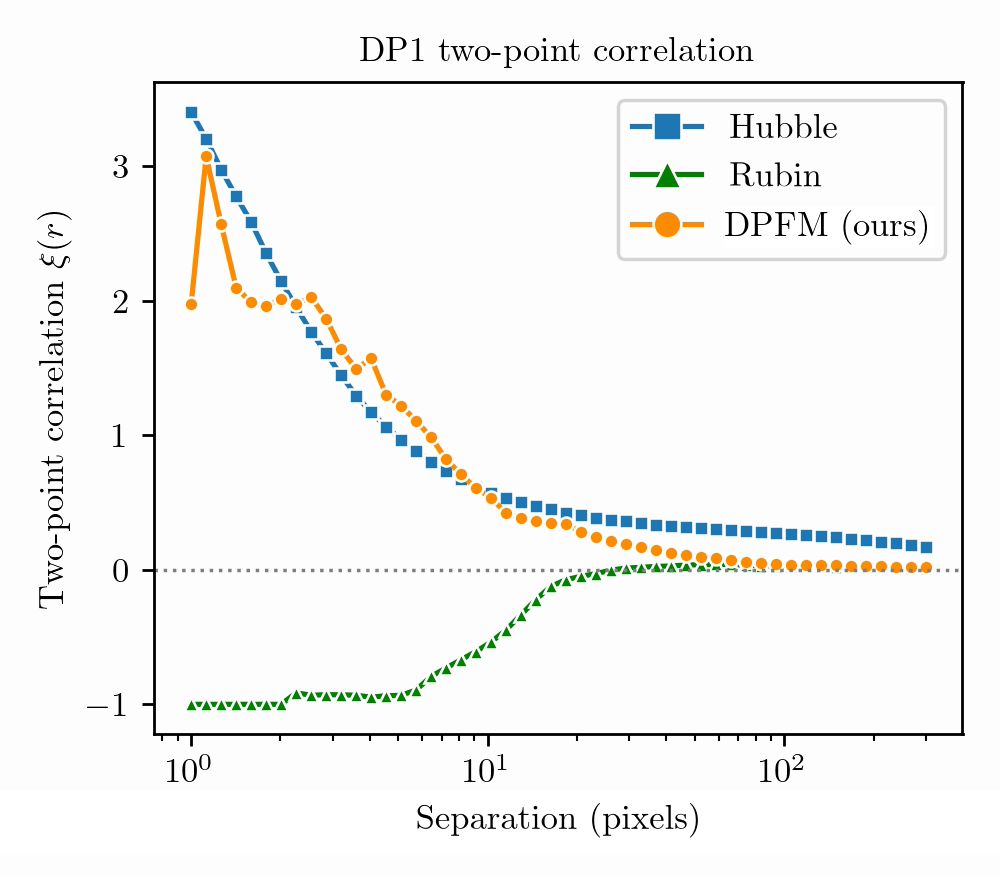}
    \caption{Two-point correlation function $\xi(r)$ on the DC2 (left) and DP1 (right) datasets as a function of source separation, for the Rubin pipeline, DPFM, and either simulation truth (DC2) or the Hubble reference (DP1).}
    \label{fig:dc2_and_dp1_two_point_correlations}
\end{figure}

Calibrated per-source marginals do not guarantee that the spatial pattern of detections reproduces the truth. We check this with the two-point correlation function $\xi(r)$, the excess probability, relative to a homogeneous Poisson process, of finding a pair of sources separated by $r$ pixels. For area elements $dA_1$ and $dA_2$ at separation $r$ in a field of mean density $\bar{n}$,
\begin{equation*}
    dP = \bar{n}^2 [1 + \xi(r)]\, dA_1\, dA_2.
\end{equation*}
Positive $\xi(r)$ indicates clustering. Because $\xi(r)$ depends jointly on the number and arrangement of sources, a catalog can recover the correct number of sources and still get $\xi(r)$ wrong, making this a stringent test. Recovering $\xi(r)$ is also a scientific end in itself: it is a primary probe of large-scale structure \citep{levine2025galaxy}.

\Cref{fig:dc2_and_dp1_two_point_correlations} (left) shows $\xi(r)$ for three DC2 catalogs: the ground truth, posterior samples from the proposed method, and the Rubin pipeline's catalog. The proposed method tracks the truth at every separation. The Rubin catalog gives $\xi(r) \approx -1$ for $r < 5$ pixels, meaning it reports almost no source pairs closer together than the PSF width, as expected if its deterministic deblender merges closely separated sources into a single detection.

\subsection{Runtime}
\label{sec:runtime}

\begin{table}[t]
\centering
\begin{tabular}{@{}cccccc@{}}
\toprule
\shortstack{Image\\Size} &
\shortstack{PCAT\\{[}Feder2020{]}} &
\shortstack{AstroPhot\\{[}Stone2023{]}} &
\shortstack{Ind-NPE\\{[}Liu2023{]}} &
\shortstack{AR-NPE\\{[}Regier2026{]}} &
\shortstack{DPFM\\(ours)} \\
\midrule
32$\times$32 & 22.9 & 24.5 & 0.006 & 0.010 & 0.142 \\
128$\times$128 & 37.0 & 85.5 & 0.006 & 0.011 & 0.144 \\
512$\times$512 & 842.1 & 5339.8 & 0.013 & 0.086 & 0.197 \\
2048$\times$2048 & 13474$^\dagger$ & 85437$^\dagger$ & 0.241 & 1.371 & 2.306 \\
\bottomrule
\end{tabular}
\caption{Runtime (in seconds) on images of varying size, with each method drawing three posterior samples. All timings use six-band image cutouts at realistic source densities and a machine with one NVIDIA RTX 5090 GPU and an AMD Ryzen 9 9900X 12-core processor (warm, mean of three runs). $^\dagger$Projected from the earlier measurements; the MCMC methods were not run at this scale.}
\label{tab:inference_runtime_by_image_size}
\end{table}

Amortized inference pays its main cost once during training, whereas MCMC incurs its full cost for every image.
At survey scale, training cost is small compared to inference cost.
\Cref{tab:inference_runtime_by_image_size} shows that DPFM is thousands of times faster than the MCMC comparators at survey-scale inference.
For the MCMC comparators, runtime cannot be considered in isolation from performance, because MCMC is asymptotically exact. The settings timed here are the same ones used in our
performance comparisons (\Cref{sec:semi_synthetic_dp1_calibration}), where the MCMC compute budget was at the limit of what was practical, and where MCMC nevertheless did not produce a
well-calibrated posterior.

AstroPhot and the neural methods are GPU-accelerated, while PCAT
runs on the CPU. Ind-NPE computes its per-image distributional parameters in one forward pass and
draws all samples from them at negligible cost. AR-NPE computes image features
once per image and generates samples through a four-step autoregressive sequence.
DPFM computes image features once per image and integrates
its probability flow over $512\times512$ windows matching its training crop, with 10 detection-phase steps and 25 measurement-phase steps.
For the $2048\times2048$-pixel image, AR-NPE and DPFM process nonoverlapping $512\times512$ chunks (a whole-image pass exceeds GPU memory for AR-NPE), while Ind-NPE processes the image in one forward pass.

Although more computationally intensive than earlier NPE-based cataloging methods, requiring tens of forward passes per posterior sample,
DPFM largely retains the survey-scale capabilities of the NPE-based methods.
The LSST full data release will cover roughly 18{,}000 square degrees at a pixel scale of 0.2 arcseconds \citep{ivezic2019lsst}, equating to $6 \times 10^{12}$ pixels.
At 2.3 seconds per $2048\times2048$ image, processing the full coadd with DPFM would take roughly 40 GPU-days. Disjoint sky regions can be processed in parallel without communication.

\section{Application to DP1}
\label{sec:dp1_application}

We train our inference networks on DC2 (\Cref{sec:simulated_datasets}) and apply them to DP1 (\Cref{sec:dp1_dataset}) to infer a catalog.
Although real DP1 imagery lacks a ground-truth catalog, a small subset of the DP1 footprint has also been imaged by the Hubble Space Telescope, which has far higher angular resolution than ground-based DP1 imagery. We use catalogs derived from these images \citep{whitmore2016hubble,haussler2007gems} as independent reference data and refer to them collectively as the Hubble reference. All DP1 results shown here are for ECDFS.

\subsection{Detection}
\label{sec:dp1_detection}

\begin{figure*}[tb]
\centering
\begin{minipage}[t]{0.62\textwidth}
    \vspace{0pt}
    \centering
    \small
    \setlength{\tabcolsep}{2.5pt}
    \begin{tabular}{c r rr rr}
    \toprule
    & & \multicolumn{2}{c}{Completeness (\%)}
    & \multicolumn{2}{c}{Purity (\%)} \\
    Magnitude bin & $N$ & Rubin & DPFM & Rubin & DPFM \\
    \midrule
    $20$--$21$   & 599    & 67.6 & 65.2 & 81.8 & 80.1 \\
    $21$--$22$   & 986    & 77.7 & 76.0 & 78.7 & 80.3 \\
    $22$--$23$   & 2,270  & 75.8 & 75.5 & 77.5 & 81.0 \\
    $23$--$24$   & 5,214  & 72.8 & 74.9 & 69.9 & 73.3 \\
    $24$--$25$   & 10,949 & 70.2 & 74.9 & 55.0 & 59.2 \\
    $25$--$26$   & 14,874 & 67.7 & 76.4 & 31.1 & 34.8 \\
    \midrule
    $20$--$26$ & 34,892 & 70.1 & 75.4 & 47.7 & 51.1 \\
    \bottomrule
    \end{tabular}
    \captionof{table}{Detection completeness and purity with the Hubble reference acting as truth, within the region where Hubble and DP1 overlap. $N$ is the number of Hubble sources in each F606W bin and hence the completeness denominator.}
    \label{tab:dp1_hubble_detection_quality}
\end{minipage}
\hfill
\begin{minipage}[t]{0.34\textwidth}
    \vspace{0pt}
    \centering
    \includegraphics[width=.9\linewidth]{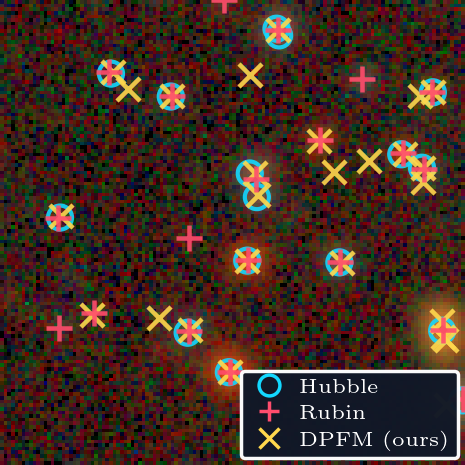}
    \captionof{figure}{Detections from DPFM, the Rubin pipeline, and Hubble overlaid on a DP1 image cutout.}
    \label{fig:dp1_detection_comparisons}
\end{minipage}
\end{figure*}

Of the Hubble sources with F606W-band magnitudes in $[20, 26]$, DPFM recovers 75.4\% to Rubin's 70.1\%, at a purity of 51.1\% against Rubin's 47.7\%. The two are close at bright magnitudes and separate at the faint end, where DPFM recovers 76.4\% of sources to Rubin's 67.7\% (\Cref{tab:dp1_hubble_detection_quality}); Appendix D.1 gives further details of the cross-catalog matching procedure.
\Cref{fig:dp1_detection_comparisons} shows detection results for an illustrative region from the DP1 dataset.

\Cref{fig:dc2_and_dp1_two_point_correlations} (right) shows the corresponding two-point correlation function $\xi(r)$ estimated from the DP1 field using posterior samples produced by DPFM, the catalog reported by the Rubin pipeline, and the Hubble reference. Both DPFM and the Hubble reference recover a strong positive clustering signal at small separations, and the two share a similar shape. Each starts at a high value and decays steadily toward zero by a few tens of pixels of separation. In contrast, the Rubin pipeline does not recover the small-scale clustering signal, reporting almost no source pairs at small separations.

\subsection{Flux estimation for isolated sources}
\label{sec:dp1_isolated_flux_estimation}

We compare each method's DP1 $z$-band point estimate with the matched Hubble F850LP magnitude on $20{,}090$ isolated sources. Appendix D.2 gives the sample-construction details and shows the photometry scatterplots. DPFM's point estimate is the magnitude of its mean flux over 20 posterior draws. The median residual (method minus Hubble) is $+0.085$ magnitudes for DPFM and $+0.004$ for Rubin. The scatter of these residuals, taken as half the difference between their 84th and 16th percentiles, is $0.245$ magnitudes for DPFM and $0.221$ for Rubin. The Pearson correlation with the Hubble magnitudes is 0.975 for DPFM and 0.959 for the Rubin pipeline.

\subsection{Flux estimation for blended sources}
\label{sec:dp1_blended_flux_estimation}

\begin{figure*}[t]
\centering
\begin{minipage}[t]{0.45\textwidth}
    \centering
    \includegraphics[height=2in]{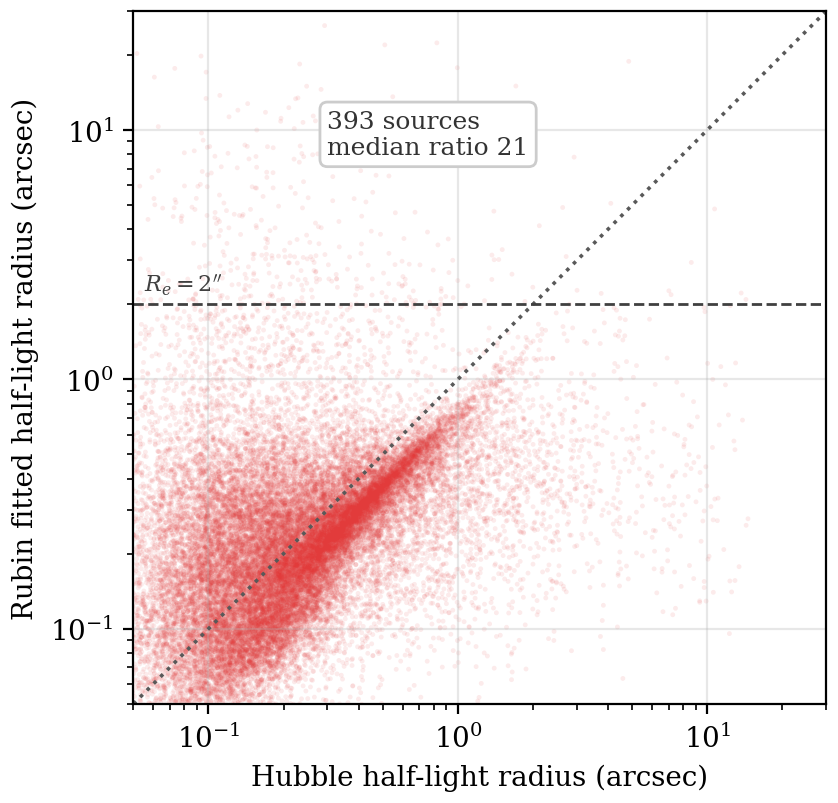}
    \caption{Rubin's half-light radius against the Hubble reference, for $33{,}121$ matched galaxies. A significant fraction of the Rubin fits lie far above the identity line.}
    \label{fig:rubin_vs_gems_half_light_radius}
\end{minipage}
\hspace{2em}
\begin{minipage}[t]{0.45\textwidth}
    \centering
    \includegraphics[height=2in]{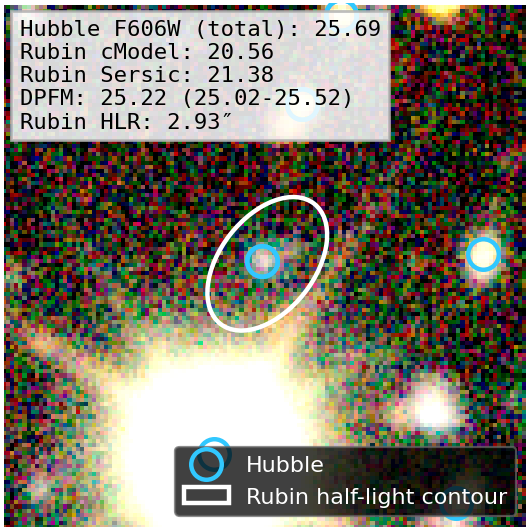}
    \caption{A representative oversized Rubin fit on a $25''$ cutout. The panel reports the magnitude for the Hubble, Rubin, and DPFM catalogs. A bright neighbor is in the lower left.}
    \label{fig:dp1_blended_flux_example}
\end{minipage}
\end{figure*}

The Rubin catalog contains a significant population of sources with severely overestimated fluxes, whereas DPFM does not.
\Cref{fig:dp1_blended_flux_example} is a representative example. Rubin's cModel and S\'ersic estimates correspond to fluxes approximately $113$ and $53$ times the Hubble flux, respectively, whereas the DPFM posterior median is $1.5$ times the Hubble flux. Appendix D.3 provides 12 additional examples.

These sources with overestimated fluxes also appear to have overestimated sizes (half-light radii).
Of the $401{,}079$ Rubin-detected sources in ECDFS, $11{,}688$ (2.9\%) have a fitted half-light radius larger than $2''$, $1{,}775$ larger than $5''$, and the largest is $29.4''$.
The oversized fits are concentrated near bright neighbors: $36\%$ lie within $15''$ of a source brighter than $r=20$, compared with $15\%$ of magnitude-matched compact controls.
Hubble imaging indicates that these fits do not generally correspond to galaxies of comparable angular extent. Matching the catalogs within $0.4''$ gives $33{,}121$ pairs; among the $393$ for which Rubin reports a half-light radius above $2''$, the median Hubble half-light radius is $0.18''$ (\Cref{fig:rubin_vs_gems_half_light_radius}). The size failures coincide with large photometric errors: $45\%$ of the oversized fits disagree with the Hubble magnitude by more than one magnitude, compared with $1.4\%$ of fits below $2''$.

DP1 is an early release, so it is possible these errors will be corrected in a future release, but Rubin's deblender and cModel photometry both predate it: the Hyper Suprime-Cam documentation reports the same population at a prevalence of a few percent \citep{hsc2021knownproblems}, and independent work finds that the pipeline's cModel sizes are unreliable at faint magnitudes \citep{huang2018characterization}. One possible reason for these failures lies in how the fit is computed rather than in the model being fitted. The least-squares objective for a PSF-convolved galaxy model is convex in a source's amplitude but not in its shape, so the fit returned can depend on where the optimizer starts. When we reimplement the documented cModel procedure \citep{bosch2018hsc} on one blended source, four starting radii give fitted half-light radii spanning more than an order of magnitude and cModel fluxes differing by nearly a factor of four, while the total $\chi^2$ is essentially unchanged (Appendix D.3.1).

DPFM performs no such per-image optimization: it samples from a posterior learned in advance. For an NPE objective of this kind, global convergence has been proven for a single-hidden-layer network in the infinite-width limit, and finite-width experiments suggest the result remains informative \citep{jacot2018neural,mcnamara2024globally}. DPFM does not precisely satisfy these conditions, so we present the optimization contrast here simply as a hypothesis.

\subsection{Calibration of flux uncertainty}
\label{sec:dp1_uncertainty_calibration}

\begin{figure}[tb]
\centering
\includegraphics[width=0.76\linewidth]{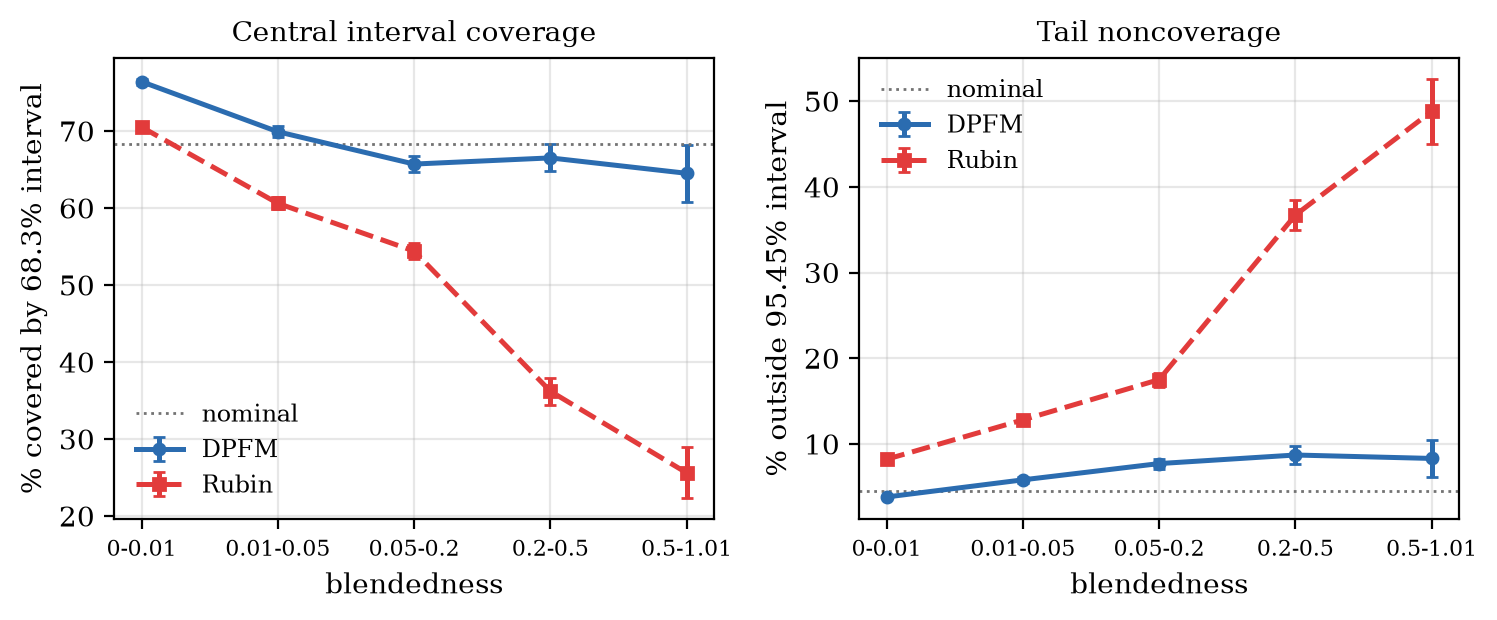}
\caption{Coverage against the Hubble reference on one common sample. The panels show the fraction inside each method's $68.3\%$ interval (left) and outside its $95.45\%$ interval (right), against Rubin blendedness; dotted lines are nominal coverage and error bars are binomial standard errors. DPFM intervals are based on posterior draws, while Rubin intervals are based on a normal distribution parameterized by the reported cModel standard error.}
\label{fig:dp1_magnitude_uncertainty_vs_blendedness}
\end{figure}

The Rubin pipeline produces a point-estimate catalog, but it also attaches an estimated standard error to each cModel flux. That error propagates the coadd's per-pixel variance estimates through the best-fit light profile; it does not propagate uncertainty in the fitted ellipse, much less uncertainty over deblendings or source configurations \citep{rubin2025lsst}.

\Cref{fig:dp1_magnitude_uncertainty_vs_blendedness} compares the empirical coverage of each method's confidence or credible intervals against the Hubble reference as blending increases. The comparison uses the same sources for both methods, binned by the blendedness reported by the Rubin pipeline. Appendix D.4 describes the sample construction and how we treat filter mismatch. For DPFM, the intervals come from its posterior draws; for Rubin, from its point estimate and standard error. In both cases, we add the estimated filter-mismatch scatter before forming the interval.

Rubin's $68.3\%$ intervals cover $70.5\%$ of the least blended sources but only $25.6\%$ of the most blended, falling monotonically in between; the fraction of sources outside its $95.45\%$ intervals rises from $8.2\%$ to $48.8\%$ over the same range. DPFM's intervals hold up better, covering $76.4\%$ of the least blended sources and $64.5$--$69.9\%$ elsewhere, with $3.8\%$ and then $5.8$--$8.7\%$ of sources outside the wider intervals. DPFM's intervals therefore slightly under-cover where blending is heaviest; Rubin's intervals fail there far more often.

\subsection{Color-magnitude diagrams}
\label{sec:dp1_color_magnitude}

Color-magnitude diagrams plot a color (the difference between a source's magnitudes in two bands) against a single band's magnitude for a sample of light sources. They can reveal the multiband photometric structure of astronomical populations present in a field.

\begin{figure*}[tb]
\centering
\begin{minipage}[t]{0.74\textwidth}\vspace{0pt}
    \centering
    \includegraphics[width=\linewidth]{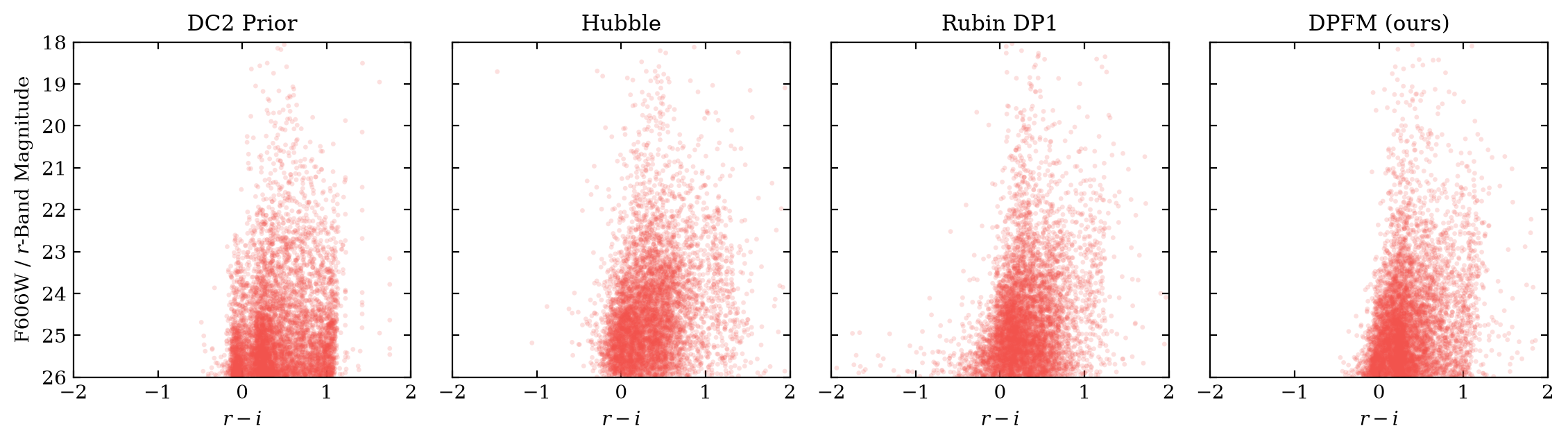}
    \caption{Color-magnitude diagrams over the $0.169$-square-degree overlap of the Hubble and DP1 footprints. Left to right: the DC2 prior (on which DPFM is trained), the Hubble reference, the Rubin catalog, and the DPFM posterior. Each panel shows $6{,}000$ randomly sampled sources.}
    \label{fig:dp1_hubble_overlap_color_magnitude_comparison}
\end{minipage}
\hfill
\begin{minipage}[t]{0.2328\textwidth}\vspace{0pt}
    \centering
    \includegraphics[width=0.82\linewidth]{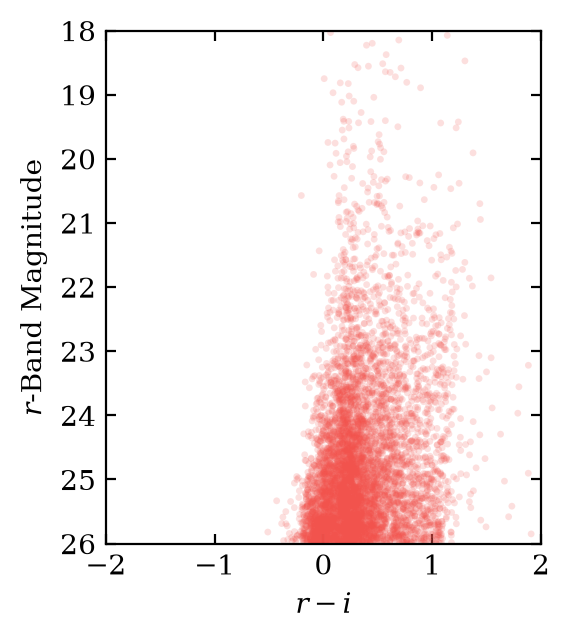}
    \caption{The same diagram from our catalog of all of ECDFS, 11 times the area.}
    \label{fig:dpfm_ecdfs_color_magnitude}
\end{minipage}
\end{figure*}

\Cref{fig:dp1_hubble_overlap_color_magnitude_comparison} shows color-magnitude diagrams for the ECDFS region imaged by Hubble. Hubble measures color through different filters than DP1, so we transform its points onto DP1's system before comparison. (Appendix D.5 describes the transformation.)
Rubin and DPFM recover similar central loci, with catalog-wide median colors differing by $0.003$ magnitudes in the $23 < r < 24$ range and $0.018$ in the $24 < r < 25$ range. In the fainter bin, the DC2 prior on which our model was trained is $0.102$ magnitudes redder than the Hubble reference, whereas DPFM is only $0.014$ magnitudes bluer and Rubin is $0.033$ magnitudes bluer. DPFM therefore sits far from its training prior and close to the reference, suggesting it responds to the input DP1 images rather than reproducing the prior.

The Rubin catalog has a small population of extremely blue sources. Within the displayed limits, $84$ of its $35{,}783$ sources ($0.235\%$) have $r-i < -1$, against $7$ of $24{,}257$ in the Hubble reference ($0.029\%$) and none among $149{,}176$ DC2 or $43{,}942$ DPFM sources.

\Cref{fig:dpfm_ecdfs_color_magnitude} extends the color-magnitude diagram of \Cref{fig:dp1_hubble_overlap_color_magnitude_comparison} to all of ECDFS, an area 11 times larger. The same locus appears. The breadth of the distribution suggests a mixed extragalactic field population spanning a range of redshifts, stellar populations, and dust content. The increasing scatter at faint magnitudes likely reflects both intrinsic diversity and greater photometric uncertainty near the survey limit.

\section{Discussion}
\label{sec:discussion}

DPFM achieves a level of fidelity and calibration unattainable with previously published probabilistic approaches to astronomical cataloging.
On simulated data, DPFM's posterior approximation is nearly indistinguishable from the exact posterior for the posterior marginals we considered, whereas each of the four probabilistic comparators was miscalibrated on at least one quantity we examined.
DPFM also assigns a higher held-out log-likelihood to simulated ground-truth catalogs than the NPE comparators do, though their claims to approximating the posterior rest on maximizing this quantity.
On real DP1 data, DPFM's credible intervals remained close to nominal coverage against an independent Hubble reference, even for the most blended sources, whereas the Rubin pipeline's intervals covered the reference only about a quarter of the time.
DPFM's computational efficiency enables it to operate at survey scale.

DPFM's gains over MCMC may be attributed to the mixing challenges faced by MCMC samplers. One of our two MCMC-based comparators is asymptotically exact, but it did not converge within a practical computational budget. The likelihood can be highly nonconvex for detection problems, and MCMC must traverse that surface to sample. DPFM's gains over the NPE comparators, on the other hand, can be explained by the flexibility of the flow-matching posterior approximation. In the NPE approaches, while the encoder network can have many parameters to map images to distributional parameters, the per-image distribution has a fairly simple parametric structure. The method of \citet{regier2026neural} overcomes one limitation of \citet{liu2023variational}, namely, the independence assumption between nearby source detections, while otherwise retaining a simple probabilistic structure.

DPFM's gains over SPFM, an ablation of it, demonstrate the importance of decoupling detection and measurement. The SPFM training objective devotes much of its gradient signal to producing null measurements for nonexistent objects. In contrast, DPFM splits inference into two phases: first, detection; second, measurement. Because DPFM performs inference in two phases, the measurement network is trained to produce only non-null values. This detect-then-measure approach to flow matching may be relevant to other image analysis problems, from detecting pedestrians in a crowd \citep{zhang2018occlusion} to inventorying forests from aerial imagery \citep{weinstein2021remote}, extending probabilistic reasoning to additional domains.

For LSST imagery, the primary alternative to DPFM is the non-probabilistic Rubin pipeline \citep{bosch2018overview,rubin2025lsst}, which adheres to an algorithmic approach that has been used for decades by predecessor surveys \citep{bertin1996sextractor,lupton2001sdss,morganson2018dark}.
Our results show some limitations of this approach.
The Rubin pipeline did not recover the small-scale spatial clustering signal, reporting almost no pairs of sources separated by less than about an arcsecond.
The Rubin pipeline reported unrealistically large sizes and correspondingly overestimated fluxes for about 3\% of sources.
The Rubin pipeline's flux uncertainties could not be used to produce calibrated error bounds in the presence of blending.
For isolated sources, the two methods' photometry is comparable; differences appear where sources overlap.
Workarounds can help, such as discarding blended objects, but they can also introduce further selection effects that are difficult to model. For instance, blend recognition may depend on the galaxies' relative orientation; in that case, discarding blends would select on shape, the quantity weak lensing sets out to measure.
In DPFM's Bayesian formulation, blending requires no special case: it arises from a forward model in which source positions have only limited dependence.

\paragraph*{Limitations and future work}
The chief limitation of the proposed method is its dependence on the simulator, which may be misspecified in two ways. First, DC2's models of the PSF, the sensor's response to bright sources, and detector artifacts such as cosmic rays and bleed trails all likely differ from those of the real instrument. Second, the prior over catalogs may misspecify the star and galaxy populations and the distribution of sources across the sky, which depends on cosmological parameters.

Existing NPE robustness techniques may help cope with the limitations of the instrument model \citep{huang2023learning,dellaporta2022robust,ward2022robust,cannon2022investigating}. A misspecified prior is a different matter: the prior here is an object of inference, not a nuisance to be made robust against. It includes the cosmological parameters LSST was built to measure, and fixing them would presuppose the answer. Therefore, training on DC2 biases the posterior toward the fiducial cosmology assumed by DC2. Nevertheless, our results suggest the data are influential: on real data, the recovered color distribution departs from the training prior toward the Hubble reference, and the recovered clustering tracks the Hubble reference as well.

More thorough removal of that prior bias could be achieved through hierarchical inference, jointly inferring the cosmological parameters with the per-image catalogs \citep{rodrigues2021hnpe}. An expectation-maximization (EM) scheme is one way to carry out this joint inference. Recent work has developed EM schemes that use neural posterior estimation for the E-step \citep{hosseintabar2025diffemlearningcorrupteddata,modi2026generative}. DPFM could potentially infer catalogs under the current cosmological parameters, and the M-step could update those parameters using the resulting posteriors, propagating detection and measurement uncertainties through to cosmological parameters.

\section*{Significance Statement}
Applied to the first public images from the NSF-DOE Vera C. Rubin Observatory, Dual-Phase Flow Matching (DPFM) recovers galaxy brightnesses and the small-scale galaxy clustering signal that the observatory's standard, deterministic catalog gets wrong: errors that would otherwise propagate, unquantified, into downstream analyses such as photometric redshifts and galaxy number counts.
Because galaxies in deep images overlap, the source configuration in each image is itself uncertain.
DPFM represents this as a probability distribution rather than a single answer, and it is thousands of times faster than Markov chain Monte Carlo at survey scales.
Detecting and measuring an unknown number of objects is a common problem in spatial statistics, so the same construction may extend fast, calibrated inference to that broad class of problems.

\section*{Acknowledgments}
This material is based on work supported by the National Science Foundation under Grant No.~2209720 and the U.S. Department of Energy, Office of Science, Office of High Energy Physics under Award Number DE-SC0023714.

\bibliographystyle{imsart-nameyear}
\bibliography{references}

\clearpage
\pdfbookmark[0]{Appendices}{appendices}
\makeatletter
{\vglue\title@skip\title@style\title@size\title@case{Appendices for ``Dual-Phase Flow Matching for Probabilistic Source Detection and Measurement in Astronomical Images''}\par}
\makeatother
\suppressfloats[t]

\appendix

\section{Implementation details}
\label{app:implementation_details}

This appendix provides details of the network architecture, training procedure, and sampling procedure of Dual-Phase Flow Matching (DPFM).
\Cref{tab:dpfm_hyperparameters} gives network hyperparameter settings.
\Cref{alg:dpfm_training,alg:dpfm_posterior_sampling} provide pseudocode.

\begin{table}[t]
\centering
\begin{tabular}{@{}ll@{}}
\toprule
Setting & Value \\ \midrule
\multicolumn{2}{@{}l}{\emph{Detection network}} \\
Backbone & UNet, four resolution levels \\
Base channel width & 16 (doubling to 256) \\
Convolution kernel & $3 \times 3$ \\
Residual block & three units, inter-unit and block residuals \\
Normalization/activation & group norm, eight groups / GELU \\
Dropout / stochastic depth & 0.1 / 0.1 \\
Image embedding & $1 \times 1$ convolution, 32 channels \\
State embedding & two $1 \times 1$ convolutions, 64 channels \\
Time embedding & two-layer network, 64 dimensions; FiLM per block \\
Parameters & $\approx 3.1$ million \\ \midrule
\multicolumn{2}{@{}l}{\emph{Measurement network}} \\
Input & $25 \times 25 \times 6$-pixel image cutouts \\
Backbone & convolutional attention \\
Heads & three (type, position, flux), three attention blocks each \\
Parameters & $\approx 3.1$ million ($0.9$M encoder, $0.7$M per head) \\ \midrule
\multicolumn{2}{@{}l}{\emph{Training}} \\
Optimizer & Adam, learning rate $10^{-3}$ \\
Learning rate schedule & cosine annealing to zero \\
Gradient-norm clip & 0.5 \\
Precision & \texttt{bfloat16} mixed precision \\
Epochs & 20 \\
Batch size & 16 random crops of one image, each $512\times{}512\times{}6$ pixels \\
Training / validation images & $4{,}000$ / $400$ \\
Image normalization & per-band inverse hyperbolic sine \\
Augmentation & axis-aligned rotations and reflections \\
GPU & NVIDIA RTX 5090 \\
CPU & AMD Ryzen 9 9900X 12-core processor \\ \midrule
\multicolumn{2}{@{}l}{\emph{Sampling}} \\
Integrator & Euler (discrete attributes); Heun, 2nd-order (continuous attributes) \\
Integration steps & 10 (detection), 25 (measurement) \\ \bottomrule
\end{tabular}
\caption{Architecture, training, and sampling settings for DPFM.}
\label{tab:dpfm_hyperparameters}
\end{table}

\subsection{Network architecture}

DPFM uses two neural networks. The detection network encodes the occupancy endpoint distribution, while the measurement network encodes the source-type endpoint distribution and the position and flux velocity fields.

\paragraph*{Detection network}
The detection network is fully convolutional, mapping an image-shaped input to one Bernoulli distribution per pixel at the native resolution of $y$. It takes the conditioning image $y$, the current occupancy pattern $x_t^{\mathrm{occ}}$, and the time $t$. Separate pointwise convolutions embed the image and state before concatenation as input to a UNet \citep{ronneberger2015u}. In a UNet, the encoder repeatedly downsamples its features, while the decoder upsamples. In addition to a bottleneck layer, skip connections connect same-size blocks in the encoder and decoder. The encoder and decoder are composed of residual blocks. A residual block has an adjustable depth and consists of convolutions, group normalization layers, and a Gaussian error linear unit (GELU) activation. The residual blocks use additive skip connections within and across blocks, and are regularized by both dropout and stochastic depth. A dense neural network supplies a time embedding to each residual block through feature-wise linear modulation (FiLM).

\paragraph*{Measurement network}
The measurement network processes a cutout around each detected source. Specifically, for each pixel with $[x_1^{\mathrm{occ}}]^i = 1$, its input is a multiband image stamp. A convolutional attention network maps each stamp to a feature vector. We again use FiLM for time conditioning.

Three structurally identical heads, for source type, position, and flux, follow the shared backbone. Each head embeds its current state together with the occupancy pattern and the other heads' current states, and fuses the result with the shared image features. Neighbor-attention blocks then exchange information among nearby detected sources using a learned relative-offset bias and per-block FiLM time conditioning.

\paragraph*{Velocity field representations}
Both networks model discrete attributes (occupancy and source type), and the measurement network models two continuous attributes (sub-pixel position and flux). The discrete heads use endpoint prediction, outputting a Bernoulli success parameter as a logit. The continuous heads output velocities $v_t^\theta$, which include two channels for position and $B$ channels for standardized log-flux. Fluxes span several orders of magnitude, which motivates the use of a standardized log-flux representation.

\begin{algorithm}[t]
\caption{Training DPFM.}
\label{alg:dpfm_training}
\begin{algorithmic}[1]
\Require simulator $p(x, y)$, step size $\eta$
\State Initialize the detection network
\Repeat
    \State Sample $(x_1, y) \sim p(x, y)$
    \State Sample $t \sim \mathrm{Uniform}[0, 1]$
    \State Sample $x_0^{\mathrm{occ}}$ uniform on $\{0,1\}^{H \times W}$
    \State Sample $x_t^{\mathrm{occ}}$ via \Cref{eq:discrete_interpolation} in the main text
    \State Compute $L_\mathrm{detect}$ via \Cref{eq:detection_loss} in the main text
    \State Update the detection network to minimize $L_\mathrm{detect}$ with step size $\eta$
\Until{convergence}
\State Initialize the measurement network
\Repeat
    \State Sample $(x_1, y) \sim p(x, y)$
    \State Sample one $t \sim \mathrm{Uniform}[0, 1]$ for all measurement heads
    \State Sample $x_0^{\mathrm{type}}$ uniform on $\{0,1\}$ where $[x_1^{\mathrm{occ}}]^i=1$
    \State Sample $x_0^{\mathrm{pos}},x_0^{\mathrm{flux}} \sim \mathcal{N}(0,I)$ where $[x_1^{\mathrm{occ}}]^i=1$
    \State Sample $x_t^{\mathrm{type}}$ and $x_t^{\mathrm c}$ using \Cref{eq:discrete_interpolation,eq:occupied_pixel_continuous_interpolation} in the main text
    \State Compute $L_\mathrm{type},L_\mathrm{pos},L_\mathrm{flux}$ where $[x_1^{\mathrm{occ}}]^i=1$ via \Cref{eq:measurement_type_loss,eq:measurement_position_loss,eq:measurement_flux_loss} in the main text
    \State Form $L_\mathrm{measure}$ from the three weighted per-head losses
    \State Update the measurement network to minimize $L_\mathrm{measure}$ with step size $\eta$
\Until{convergence}
\State \Return the trained detection and measurement networks
\end{algorithmic}
\end{algorithm}

\subsection{Training}

\Cref{alg:dpfm_training} states the training procedure. The detection and measurement networks are trained separately; the latter jointly minimizes source-type cross-entropy and squared-error losses for position and flux using learned per-head weights. Its heads share one sampled time, condition on the ground-truth occupancy pattern $x_1^{\mathrm{occ}}$, and are evaluated jointly with the same noisy source type, position, and flux states. Each training example is a $512 \times 512$ crop of a simulated image--catalog pair $(x_1,y)$, with a discrete interpolation path for source type and continuous linear paths for position and flux.

Training pairs are produced by the simulators described in \Cref{app:semi_synthetic_dp1_experiments,app:dc2_experiments}. Training takes 80{,}000 gradient steps per network. The per-band log-flux standardization constants, $\mu_b$ and $\sigma_b$, are fixed per survey using the training catalogs. A full training run takes approximately six hours (three hours per network).
We retain the parameters from the epoch with the lowest validation loss.

\begin{algorithm}[t]
\caption{Sampling a posterior catalog.}
\label{alg:dpfm_posterior_sampling}
\begin{algorithmic}[1]
\Require image $y$, trained detection and measurement networks, step counts $N_{\mathrm{det}}$, $N_{\mathrm{meas}}$
\State Sample $x_0^{\mathrm{occ}}$ uniform on $\{0,1\}^{H \times W}$
\State $h \leftarrow 1 / N_{\mathrm{det}}$ \Comment{Detection phase: local time $t \in [0, 1]$}
\For{$k = 0, \dots, N_{\mathrm{det}}-1$}
    \State $t \leftarrow kh$
    \State Compute the detection endpoint probabilities (logits)
    \State Update $x_{(k+1)h}^{\mathrm{occ}}$ by an Euler step
\EndFor
\State $x_1^{\mathrm{occ}} \leftarrow x_{N_{\mathrm{det}} h}^{\mathrm{occ}}$ \Comment{$N_{\mathrm{det}} h = 1$}
\State Sample $x_0^{\mathrm{type}}$ uniform on $\{0,1\}$ where $[x_1^{\mathrm{occ}}]^i=1$
\State Sample $x_0^{\mathrm c}\sim\mathcal{N}(0,I)$ where $[x_1^{\mathrm{occ}}]^i=1$
\State $h \leftarrow 1 / N_{\mathrm{meas}}$ \Comment{Measurement phase: local time $t \in [0, 1]$}
\For{$k = 0, \dots, N_{\mathrm{meas}}-1$}
    \State $t \leftarrow kh$
    \State Compute the source-type endpoint probabilities (logits)
    \State Update $x_{(k+1)h}^{\mathrm{type}}$ by an Euler step
    \State Evaluate $v_t^\theta(x_t^{\mathrm m},y,x_1^{\mathrm{occ}})$
    \State Advance $x_t^{\mathrm c}$ to $x_{(k+1)h}^{\mathrm c}$ by a Heun step
\EndFor
\State Read $x_1^{\mathrm m}$ at pixels with $[x_1^{\mathrm{occ}}]^i=1$
\State \Return $(x_1^{\mathrm{occ}},x_1^{\mathrm m})$
\end{algorithmic}
\end{algorithm}

\subsection{Sampling}

A posterior catalog sample is produced in two phases (\Cref{alg:dpfm_posterior_sampling}). Conceptually, the detection phase occupies the first half of the global sampling time, and the measurement phase occupies the second. Each phase has its own local flow time $t\in[0,1]$. Each invocation of the detection network returns an occupancy endpoint as a logit for each pixel. \Cref{eq:discrete_transition_rates} in the main text maps these endpoint predictions to transition rates $Q_{t,j}^\theta$ that induce the full-state rate matrix $Q_t^\theta$. For each occupancy coordinate $j$, the Euler update forms the categorical vector $\mathbf 1\{a=x_{t,j}^{\mathrm{occ}}\}+hQ_{t,j}^\theta(a\mid x_t)$ over $a\in\{0,1\}$ and samples $x_{t+h,j}^{\mathrm{occ}}$.

Given the sampled $x_1^{\mathrm{occ}}$, each invocation of the measurement network produces position and flux velocities wherever $[x_1^{\mathrm{occ}}]^i=1$. The heads output the learned velocity field $v_t^\theta(x_t^{\mathrm m},y,x_1^{\mathrm{occ}})$ directly. A second-order Heun step advances $x_t^{\mathrm c}$ to $x_{t+h}^{\mathrm c}$ by evaluating the velocity twice: at the current state and at a trial Euler step to $t+h$. These velocities are averaged to produce the update.

\section{Experiments with the semi-synthetic DP1 dataset}
\label{app:semi_synthetic_dp1_experiments}

The semi-synthetic DP1 dataset supplies a known reference catalog for calibration experiments on DP1-like images. Real DP1 imagery does not provide one, since the Rubin DP1 catalog is itself a pipeline estimate rather than ground truth. We construct the dataset by re-rendering each DP1 field from its Rubin catalog: the rendered images carry the same source population, and the Rubin catalog that produced them is, by construction, the correct catalog for those images.

\subsection{Forward simulation}

For each DP1 field, the Rubin catalog provides source positions, magnitudes, and morphological parameters (half-light radii, S\'ersic indices, and axis ratios). We pass this catalog through AstroPhot \citep{stone2023astrophot} in three steps:
\begin{enumerate}
    \item Render each cataloged source. Stars are rendered as point sources, while galaxies are rendered as S\'ersic profiles with the specified half-light radius, S\'ersic index, and axis ratio.
    \item Convolve the resulting source image with a Moffat point-spread function, whose parameters are fitted to bright, isolated stars in the same field by Levenberg--Marquardt minimization of the sum of squared residuals.
    \item Add a correlated background noise field whose statistics are measured in the same field.
\end{enumerate}
The output is one semi-synthetic image per DP1 field, matched to the original in pixel scale, depth, and seeing. For the star-only images used to test PCAT, we replace the galaxies with stars to match the source density of the original images. We use the same PSF and correlated background noise.

\subsection{MCMC comparator configurations}

AstroPhot uses the rendered images as observations to infer the parameters of a sky model jointly within each local region. It uses the same renderer, point-spread-function fit, and correlated-Gaussian pixel-noise likelihood as the forward simulation. Its source parameterization differs: the simulation renders each galaxy as a single S\'ersic profile, whereas AstroPhot represents each extended source as a bulge-and-disk system of two S\'ersic components sharing a center. Source centers are initialized from peak finding on the image.

AstroPhot draws samples from the joint posterior of these parameters within each region using its Pyro-based, fixed-path implementation of Hamiltonian Monte Carlo (HMC). We selected the leapfrog step count and step size via a grid search: 10 leapfrog steps with a step size of 0.01 converged most rapidly while remaining numerically stable. The chain runs 200 warmup iterations and 1{,}000 sampling iterations, from which three draws are retained per region.

The PCAT probabilistic-cataloging sampler \citep{feder2020multiband}, which implements reversible-jump MCMC, draws samples from the transdimensional joint posterior over the source count and each source's position and flux. The sampler is initialized by matched-filter peak detection. The chain runs 200 warmup Metropolis steps followed by 1{,}000 sampling steps, from which three posterior draws are retained per image. At each Metropolis step, the sampler proposes one of three types of moves: a within-model perturbation of positions and fluxes, a birth/death of a single source, or a merge/split of a pair, selected with probabilities 0.5, 0.25, and 0.25, respectively. The birth/death and merge/split moves change the catalog dimension and allow the chain to explore an unknown number of sources. A per-source parsimony penalty is applied to reduce false-positive detections.

This configuration is the largest sampling budget we could afford per image: even on a small 32$\times$32 image, PCAT takes 22.9 seconds, and AstroPhot takes 24.5 seconds, already more than two orders of magnitude longer than the proposed method takes.

\section{Experiments with the DC2 Simulated Sky Survey}
\label{app:dc2_experiments}

The DC2 simulated survey complements the semi-synthetic DP1 benchmark of \Cref{app:semi_synthetic_dp1_experiments}. Semi-synthetic DP1 measures agreement with the Rubin pipeline on DP1-like images, whereas DC2 measures agreement with an exact ground-truth catalog on simulated LSST-like images.

\subsection{The DC2 simulation}
The LSST DESC Data Challenge 2 (DC2) \citep{abolfathi2021lsst} is a fully synthetic, end-to-end LSST simulation. The simulator has three stages. First, a truth catalog is constructed. A gravity-only $N$-body simulation generates the dark-matter halos and their light cones. Galaxies are assigned to those halos. Ray tracing is used to simulate lensing shear and magnification. Stars are added based on a Milky Way model, and active galactic nuclei and supernovae are added as well. Second, an image simulator renders LSST-like exposures in the six bands, $u, g, r, i, z, y$, over a $\sim 300\,\mathrm{deg}^2$ region, following a five-year baseline observing cadence and including atmospheric and instrumental effects. Third, the LSST Science Pipelines process these exposures into calibrated coadds and catalogs designed to resemble planned LSST data products. Because the input truth catalog is known exactly and is matched to the detected objects, any downstream catalog can be evaluated against it directly.

\begin{figure}[t]
    \centering
    \includegraphics[width=0.45\linewidth]{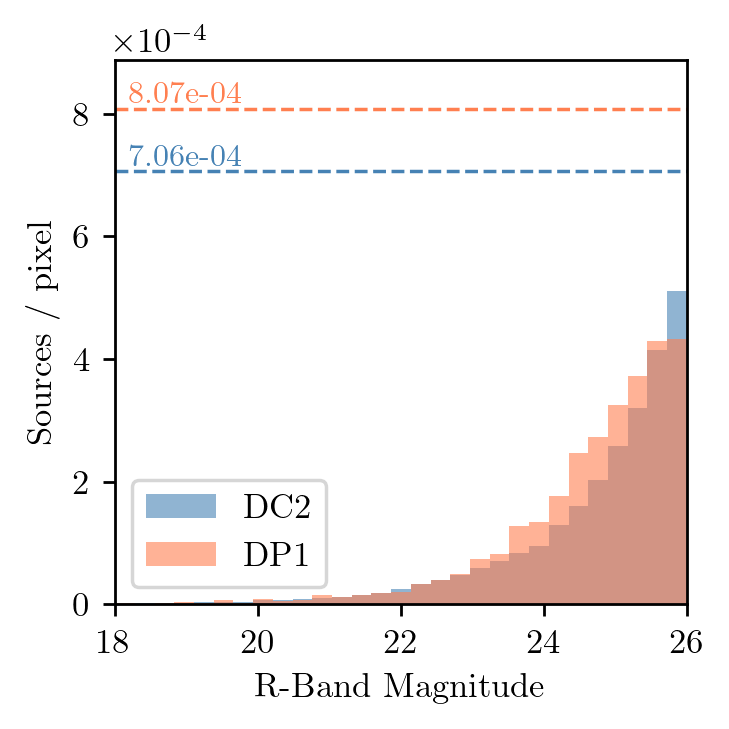}
    \caption{Per-bin source density (sources per pixel) as a function of $r$-band magnitude. Dashed lines indicate the total source density summed over all bins. DP1 results are from the Rubin pipeline, applied to DP1.}
    \label{fig:dc2_vs_dp1_source_density_by_magnitude}
\end{figure}

\Cref{fig:dc2_vs_dp1_source_density_by_magnitude} shows that the source density in DC2 agrees closely, though not exactly, with the source density Rubin identifies in DP1.

\subsection{Detection ambiguity example}
\begin{figure}[tb]
\centering
\begin{minipage}[t]{0.25\textwidth}\vspace{0pt}
  \centering
  \includegraphics[width=\linewidth]{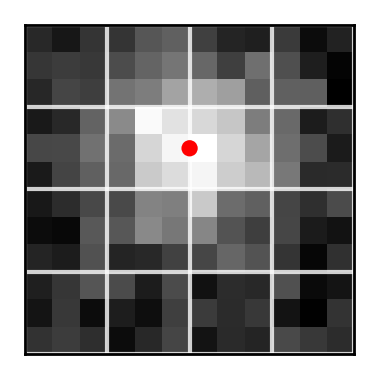}
\end{minipage}%
\begin{minipage}[t]{0.25\textwidth}\vspace{0pt}
  \centering
  \includegraphics[width=\linewidth]{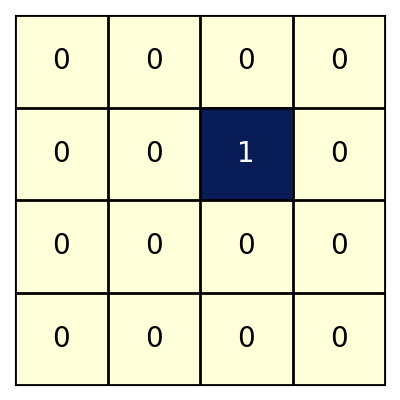}
\end{minipage} %
\begin{minipage}[t]{0.25\textwidth}\vspace{0pt}
  \centering
  \includegraphics[width=\linewidth]{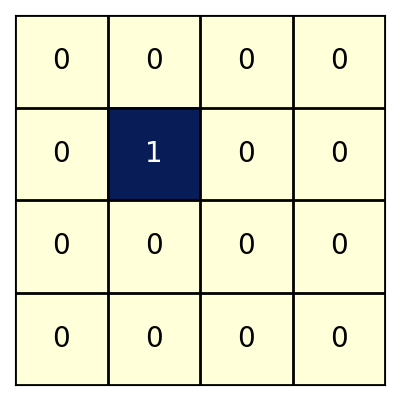}
\end{minipage} %
\caption{A galaxy on a pixel boundary. Left: an image; center and right: two detection samples.}
\label{fig:pixel_boundary_detection_examples}
\end{figure}

A source whose center sits near a pixel boundary can produce posterior samples that detect it in either of two neighboring pixels. \Cref{fig:pixel_boundary_detection_examples} shows two such samples from DPFM. In the DC2 validation set, the rate of double-detection samples is 1.97\% for DPFM, compared with 2.01\% for the AR-NPE sampler of \citet{regier2026neural}. Thus, on this boundary case, DPFM achieves essentially the same double-detection rate as explicit autoregressive sampling, without requiring an autoregressive factorization.

\subsection{Source-type classification}
\begin{figure}[tb]
    \centering
    \includegraphics[height=2in]{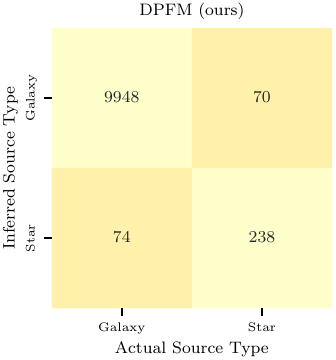}
    \hspace{1em}
    \includegraphics[height=2in]{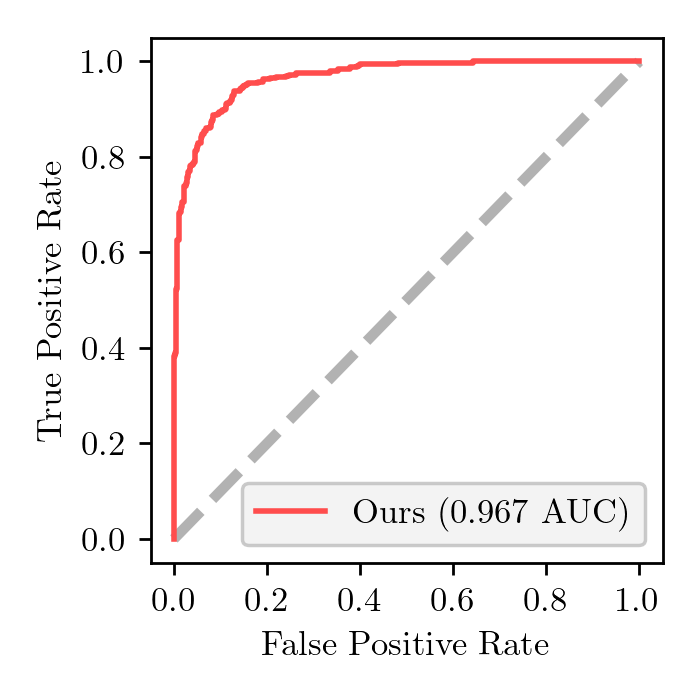}
    \caption{Source-type inference. Cell colors encode the asymmetry of each transpose pair: lighter colors indicate smaller differences, and deeper colors indicate greater asymmetry.}
    \label{fig:dpfm_star_galaxy_classification}
\end{figure}

\Cref{fig:dpfm_star_galaxy_classification} shows star-versus-galaxy classification by DPFM. The confusion matrix is dominated by its diagonal (9{,}948 galaxies and 238 stars correctly classified, against off-diagonal counts of 70 and 74); the absolute misclassification counts are similar for the two classes, though the per-class error rate is higher for the rarer stars. The receiver operating characteristic curve stays close to the upper-left corner, with an area under the curve of 0.967.

\subsection{Frequentist coverage of credible intervals}
We additionally assess frequentist coverage of credible intervals for the continuous quantities we infer, namely, sub-pixel position and per-band flux. We form highest-density credible intervals for four nominal levels. The credible intervals are conditioned on the true occupancy pattern, so that detection performance does not confound this analysis. For a perfectly calibrated posterior, with enough samples, empirical coverage matches the nominal level. Calibrated coverage is necessary but not sufficient to establish posterior calibration; credible intervals based solely on the prior would also exhibit nominal coverage.

\begin{figure}[tb]
    \centering
    \includegraphics[width=0.68\textwidth]{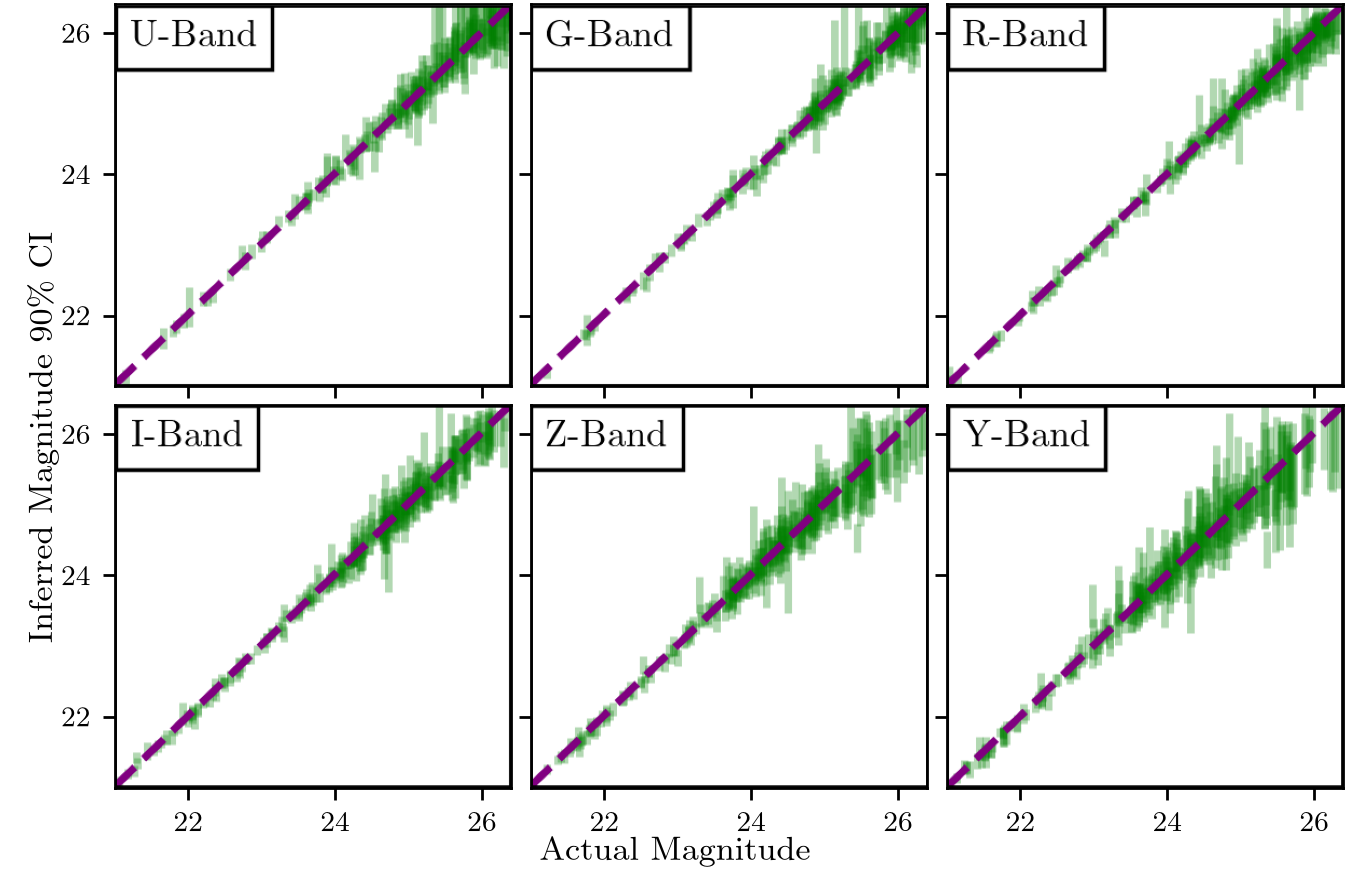}
    \caption{The highest-density 90\% credible intervals for the band-specific fluxes of light sources. The 45-degree line represents perfect estimation accuracy.}
    \label{fig:dc2_flux_90_percent_credible_intervals}
\end{figure}

Across the 32 cells in \Cref{tab:credible_interval_coverage}, DPFM is closest to nominal coverage in 27, including every position cell; AR-NPE is closest in the remaining five flux cells, and Ind-NPE in none. At the 90\% level, DPFM has the smallest coverage error for both position coordinates and every flux band except $y$. For example, its horizontal-position error is $0.011$, compared with $0.053$ for Ind-NPE and $0.044$ for AR-NPE, while its $r$-band error is $0.009$, compared with $0.047$ and $0.018$. The comparators generally over-cover, indicating intervals wider than the data warrant. \Cref{fig:dc2_flux_90_percent_credible_intervals} plots the per-source 90\% flux intervals against ground truth.

\begin{table}[tb]
    \centering
    \footnotesize
    \setlength{\tabcolsep}{4.5pt}
    \begin{tabular}{@{}c|ccc|ccc@{}}
    \toprule

    \multirow{2}{*}{\shortstack{Expected\\Coverage}}
    & \multicolumn{3}{c|}{Horizontal Position}
    & \multicolumn{3}{c}{Vertical Position} \\
    \cmidrule(l){2-7}
    & \shortstack{Ind-NPE\\{[}Liu2023{]}} & \shortstack{AR-NPE\\{[}Regier2026{]}} & \shortstack{DPFM\\(ours)}
    & \shortstack{Ind-NPE\\{[}Liu2023{]}} & \shortstack{AR-NPE\\{[}Regier2026{]}} & \shortstack{DPFM\\(ours)} \\
    \midrule
    90\% & 0.953 & 0.944 & \textbf{0.889} & 0.944 & 0.937 & \textbf{0.908} \\
    80\% & 0.878 & 0.882 & \textbf{0.801} & 0.892 & 0.879 & \textbf{0.816} \\
    70\% & 0.803 & 0.811 & \textbf{0.693} & 0.834 & 0.810 & \textbf{0.689} \\
    50\% & 0.661 & 0.644 & \textbf{0.523} & 0.658 & 0.639 & \textbf{0.505} \\
    \midrule

    & \multicolumn{3}{c|}{Flux $u$}
    & \multicolumn{3}{c}{Flux $g$} \\
    \cmidrule(l){2-7}
    90\% & 0.930 & 0.912 & \textbf{0.901} & 0.932 & 0.913 & \textbf{0.898} \\
    80\% & 0.811 & 0.813 & \textbf{0.809} & 0.837 & 0.820 & \textbf{0.802} \\
    70\% & 0.744 & 0.718 & \textbf{0.717} & 0.759 & 0.733 & \textbf{0.693} \\
    50\% & 0.552 & \textbf{0.512} & 0.519 & 0.550 & 0.533 & \textbf{0.474} \\
    \midrule

    & \multicolumn{3}{c|}{Flux $r$}
    & \multicolumn{3}{c}{Flux $i$}\\
    \cmidrule(l){2-7}
    90\% & 0.947 & 0.918 & \textbf{0.891} & 0.931 & 0.914 & \textbf{0.898} \\
    80\% & 0.854 & 0.818 & \textbf{0.809} & 0.844 & \textbf{0.815} & 0.816 \\
    70\% & 0.788 & 0.732 & \textbf{0.706} & 0.745 & 0.721 & \textbf{0.696} \\
    50\% & 0.610 & 0.523 & \textbf{0.495} & 0.542 & 0.534 & \textbf{0.529} \\
    \midrule

    & \multicolumn{3}{c|}{Flux $z$}
    & \multicolumn{3}{c}{Flux $y$} \\
    \cmidrule(l){2-7}
    90\% & 0.924 & 0.922 & \textbf{0.884} & 0.931 & \textbf{0.901} & 0.911 \\
    80\% & 0.830 & 0.812 & \textbf{0.799} & 0.842 & 0.814 & \textbf{0.809} \\
    70\% & 0.732 & \textbf{0.716} & 0.720 & 0.773 & 0.713 & \textbf{0.700} \\
    50\% & 0.527 & 0.522 & \textbf{0.519} & 0.590 & \textbf{0.503} & 0.485 \\
    \bottomrule
    \end{tabular}%
    \caption{The frequentist coverage of highest-density credible intervals. We consider four levels of coverage.}
    \label{tab:credible_interval_coverage}
\end{table}

\section{Supporting analyses for the DP1 application}
\label{app:dp1_supporting_analyses}

This appendix provides details of our numerical comparisons between DPFM and the Rubin pipeline using DP1 data, with two Hubble catalogs as references.

\subsection{Detection against the Hubble reference}
\label{app:dp1_hubble_detection_matching}

To assess detection performance, as a surrogate for ground truth, we use the Hubble Source Catalog version 3 ``detailed'' table, which contains distinct rows for different visits, detectors, and filters.
We use the F606W band to measure flux, drawing on both the MagAuto and MagAper2 columns to create our reference.
We use MagAper2 when MagAuto is missing and do not retain sources missing both measurements.
We also require sources to appear in more than one Hubble image to be included in our reference, as recommended by \citet{whitmore2016hubble}; sources appearing only once may be artifacts.

For cross-catalog matching, we compare Hubble F606W-band detections with DP1 $r$-band detections, as their filter curves are similar.
In DP1, some sources appear in multiple patches because the patches overlap; we assign each source to a unique patch.
Our comparison is restricted to the Hubble coverage of ECDFS with valid DP1 pixels.
We perform one-to-one matching into the Hubble reference catalog, with candidate pairs ordered from smallest to largest separation,
and a maximum matching distance of $0.5''$.
We also require magnitudes to match within one magnitude.
For DPFM, for stability, we perform matching with 20 separate posterior catalogs and report average performance across them; we do not take a union of DPFM detections across posterior samples, or otherwise combine multiple DPFM samples to produce a point-estimate catalog
that is better for the reported metrics, although doing so could be interesting future work.

For the completeness metric, the denominator is defined by the Hubble reference in each magnitude bin.
For the purity metric, the denominator is defined as the number of detections in each DP1 catalog, stratified by predicted magnitude.

\subsection{Isolated-source photometry against the GEMS reference}
\label{app:dp1_gems_photometry}

\begin{figure}[b]
    \centering
    \includegraphics[width=0.4\linewidth]{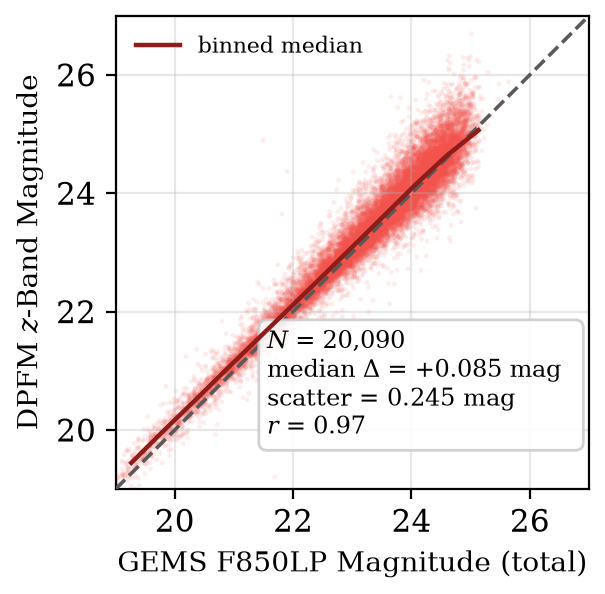}
    \hspace{1em}
    \includegraphics[width=0.4\linewidth]{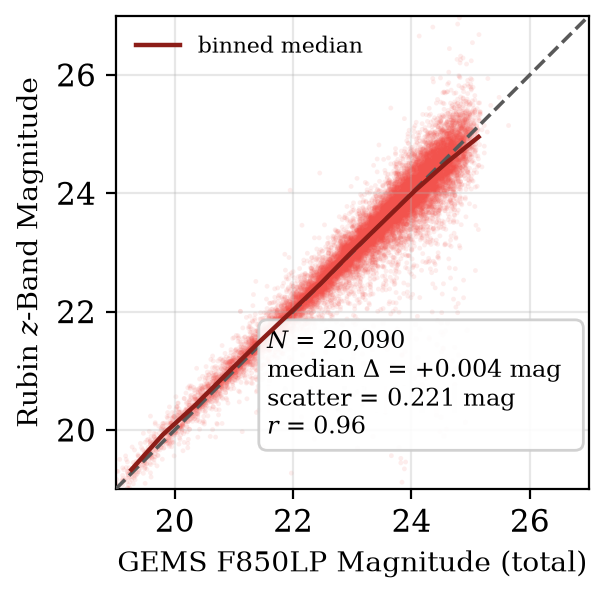}
    \caption{The magnitude of DPFM's posterior mean flux (left) and DP1 $z$ magnitudes from the Rubin pipeline (right) against GEMS F850LP total magnitudes, on the same isolated sources. The DPFM mean uses 20 posterior draws; sources detected in fewer than half are excluded. Each panel reports statistics over the full matched sample; the plotting window is restricted to $19$--$27$ magnitudes and omits 189 brighter reference sources.}
    \label{fig:rubin_and_dpfm_vs_gems_magnitudes}
\end{figure}

For isolated-source photometry, we use the GEMS Hubble reference as a proxy for ground truth.
We perform matching using a maximum radius of $0.4''$ and compare the DP1 $z$-band flux with the GEMS F850LP-band flux; these two bands have similar filter curves.
DPFM sources are retained only when detected in at least 10 of 20 posterior draws, and the DPFM flux point estimate is equal to the mean of the matched fluxes.
\Cref{fig:rubin_and_dpfm_vs_gems_magnitudes} shows sources detected by both methods.

\subsection{Blended-source photometry and oversized Rubin fits}
\label{app:dp1_blended_photometry}

\begin{figure*}[tb]
    \centering
    \includegraphics[width=0.92\textwidth]{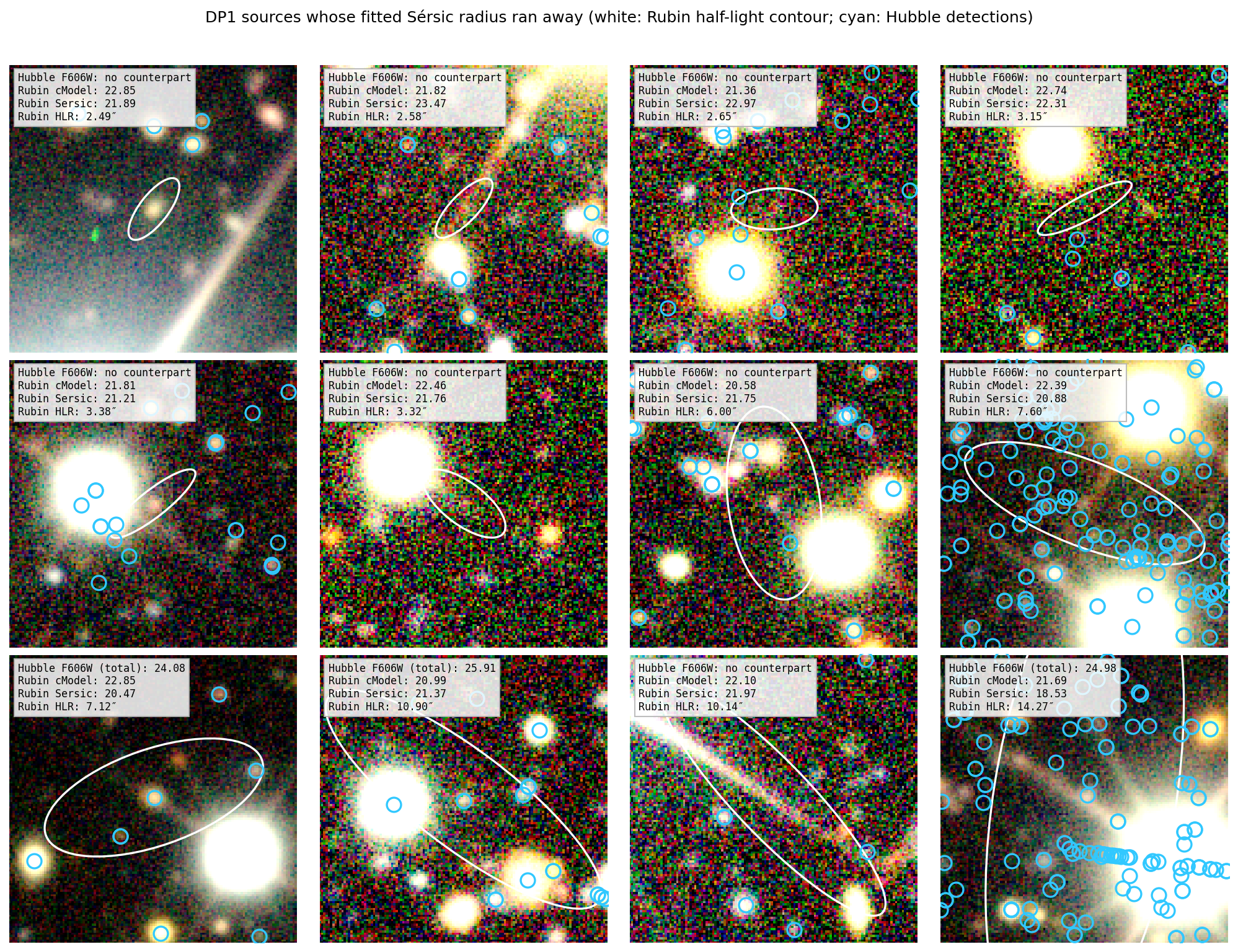}
    \caption{Twelve DP1 sources fitted with a half-light radius above $2''$, on $25''$ $r$-band cutouts. The white ellipse is the half-light contour implied by the Rubin catalog's S\'ersic parameters, and cyan circles mark Hubble detections. Each panel reports the Hubble total magnitude where a counterpart exists, along with both Rubin fits and our posterior median.}
    \label{fig:dp1_oversized_rubin_fit_gallery}
\end{figure*}

\Cref{fig:dp1_oversized_rubin_fit_gallery} contains 12 representative examples of light sources that Rubin catalogs as having a half-light radius above $2''$.

\subsubsection{Sensitivity of the cModel fit to its starting point}
\label{app:cmodel_multistart_sensitivity}

To better understand why the Rubin pipeline overestimates cModel half-light radii and
fluxes for some sources, we reimplement its documented three-stage fitting routine and
apply it to several apparent failures. Our reimplementation is a surrogate rather than
the pipeline's own code: it first fits an exponential profile, then a de Vaucouleurs
profile, and finally a nonnegative combination of the two with their shapes held fixed.
It also uses our own masking of neighboring sources and a SciPy optimizer with
independently chosen parameter bounds, so the experiment probes the behavior of the
fitting procedure without attempting to reproduce the pipeline's exact results.

\begin{figure}[tb]
    \centering
    \includegraphics[width=0.8\linewidth]{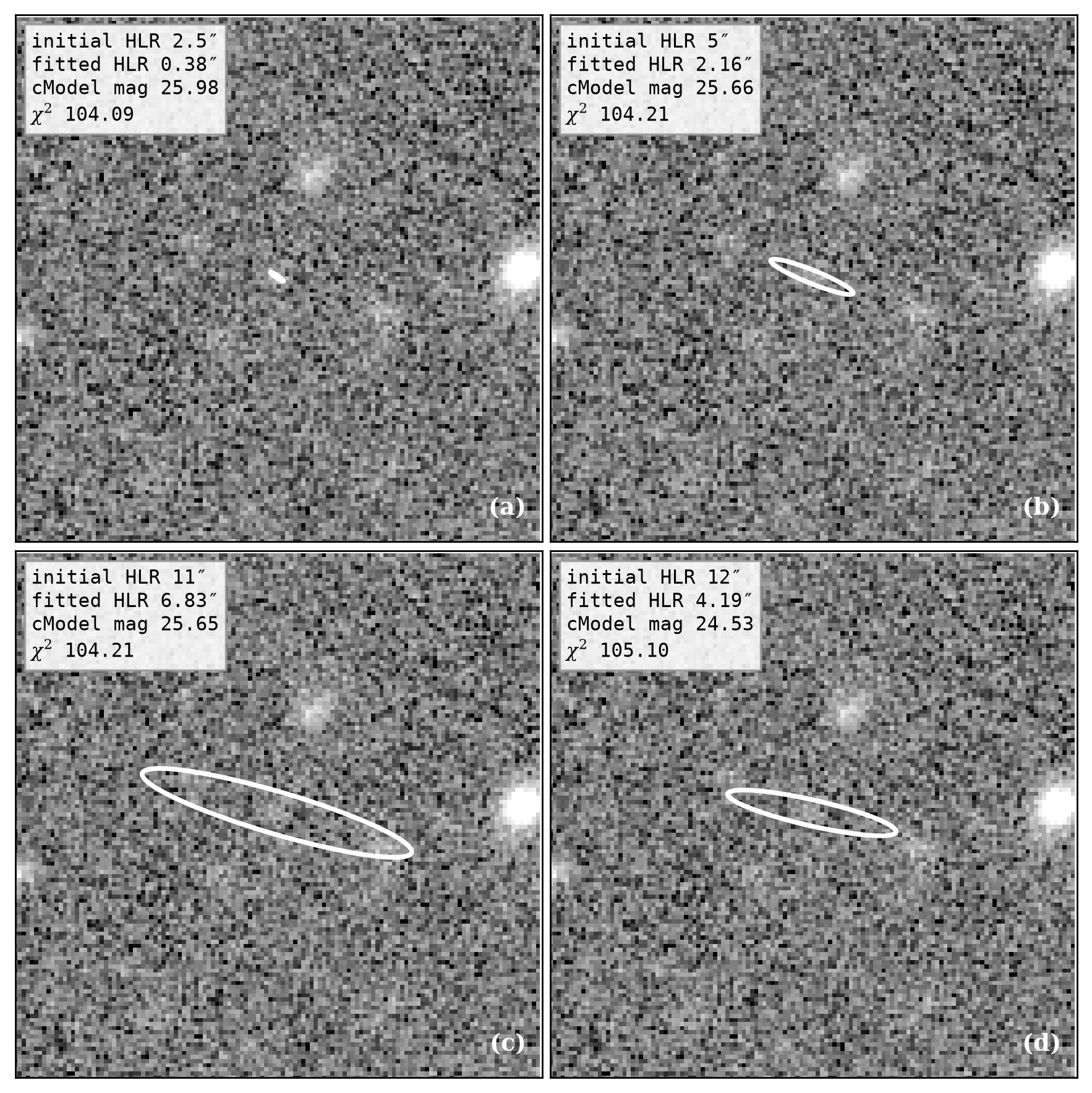}
    \caption{DP1 $r$-band coadd pixels at $(\mathrm{R.A.},\mathrm{decl.})=(52.945692^\circ,-27.621458^\circ)$, in tract 5063, patch 45. The same blended source is fitted from four starting half-light radii. Each panel shows the recovered exponential-component half-light contour in white and gives the initial radius, fitted radius, cModel magnitude, and total $\chi^2$. The ellipse and magnitude summarize different parts of the three-stage fit: the ellipse shows the exponential component's fitted radius, whereas the cModel magnitude uses the total flux of the final exponential--de Vaucouleurs combination.}
    \label{fig:dp1_cmodel_multistart_sensitivity}
\end{figure}

\Cref{fig:dp1_cmodel_multistart_sensitivity} shows a representative example of initialization sensitivity. Fitting the same
pixels from four initial half-light radii produces fitted radii ranging from $0.38''$ to $6.83''$ and cModel magnitudes ranging
from $24.53$ to $25.98$. The corresponding total $\chi^2$ values range only from $104.09$ to $105.10$ over 125 fitted pixels.
The dependence of the recovered solution on its initialization suggests a nonconvex objective with multiple local optima, where the starting radius determines the optimizer's convergence point.

\subsection{Uncertainty coverage against blendedness}
\label{app:dp1_uncertainty_coverage}

To assess uncertainty coverage, we compare both methods' DP1 $z$-band flux predictions with the GEMS F850LP-band flux estimates; these bands have similar but not identical filter curves.
We convert DP1 fluxes from nJy to AB magnitudes.
For matching, we use a $0.4''$ maximum separation and then restrict to sources detected by both methods, requiring a match in at least 10 of the 20 DPFM posterior samples.

We attempt to account for filter-curve differences using the observed
$z-\mathrm{F850LP}$ scatter among minimally blended sources.
We take half the difference between the 84th and 16th percentiles of the Rubin $z$ minus GEMS F850LP residuals ($0.186$ magnitudes) as our estimate of scatter.
For DPFM, we convolve the empirical posterior with a zero-centered normal distribution
having this standard deviation and calculate central $68.3\%$ and $95.45\%$ intervals
from the resulting distribution. For Rubin, we convert the cataloged flux error to a
magnitude error, combine it with $0.186$ magnitudes by adding their squared values, and use
one and two times the resulting error for the $68.3\%$ and $95.45\%$ intervals,
respectively.

\subsection{Color-magnitude comparison}
\label{app:dp1_color_magnitude}

To create color-magnitude diagrams that are comparable across DP1 and Hubble, we transform the Hubble F606W and F814W photometry
to estimated DP1 $r$ and $i$ photometry. This transformation is necessary because F606W and F814W are more widely separated in
effective wavelength than $r$ and $i$, so the corresponding colors are not directly comparable. We calibrate the transformation
using 718 Hubble--Rubin matches with $\mathrm{F606W}<23$, exactly one Hubble source within $1''$, and a Rubin half-light radius
below $2''$. We fit the relations by least squares, remove calibration sources with residuals exceeding $3\sigma$, and
repeat the fit, ultimately arriving at the following models:
\begin{align*}
r-i &= -0.061 + 0.739\,(\mathrm{F606W}-\mathrm{F814W}), \\
r &= \mathrm{F606W}-0.300-0.157\,(\mathrm{F606W}-\mathrm{F814W}).
\end{align*}
The residual scatter is $0.059$ magnitudes in $r - i$ color and $0.153$ magnitudes in $r$-band magnitude.

\end{document}